\documentclass[12pt,a4paper]{article}
\pdfoutput=1

\usepackage{parskip}

\usepackage[utf8]{inputenc}
\usepackage[T1]{fontenc}

\usepackage{multirow}
\usepackage{amsmath}
\usepackage{color}
\usepackage{amsfonts}
\usepackage{amssymb}
\usepackage{graphicx}
\usepackage{placeins}
\usepackage{geometry}
\usepackage{amssymb,epsfig,subfigure}
\usepackage{hyperref}
\usepackage{url}
\usepackage{comment}
\usepackage[font=footnotesize]{caption}
\usepackage{slashed}

\makeatletter
\renewcommand\section{\@startsection {section}{1}{\z@}%
                                 {-3.5ex \@plus -1ex \@minus -.2ex}
                                   {2.3ex \@plus.2ex}%
                                   {\normalfont\large\bfseries}}
\renewcommand\subsection{\@startsection{subsection}{2}{\z@}%
                                   {-3.25ex\@plus -1ex \@minus -.2ex}%
                                     {1.5ex \@plus .2ex}%
                                     {\normalfont\bfseries}}
\renewcommand\subsubsection{\@startsection{subsubsection}{3}{\z@}%
                                   {-3.25ex\@plus -1ex \@minus -.2ex}%
                                     {1.5ex \@plus .2ex}%
                                     {\normalfont\itshape}}
\makeatother

\def\pplogo{\vbox{\kern-\headheight\kern -29pt
\halign{##&##\hfil\cr&{\ppnumber}\cr\rule{0pt}{2.5ex}&\ppdate\cr}}}
\makeatletter
\def\ps@firstpage{\ps@empty \def\@oddhead{\hss\pplogo}%
  \let\@evenhead\@oddhead 
}
\def\maketitle{\par
 \begingroup
 \def\thefootnote{\fnsymbol{footnote}}
 \def\@makefnmark{\hbox{$^{\@thefnmark}$\hss}}
 \if@twocolumn
 \twocolumn[\@maketitle]
 \else \newpage
 \global\@topnum\z@ \@maketitle \fi\thispagestyle{firstpage}\@thanks
 \endgroup
 \setcounter{footnote}{0}
 \let\maketitle\relax
 \let\@maketitle\relax
 \gdef\@thanks{}\gdef\@author{}\gdef\@title{}\let\thanks\relax}
\makeatother

\numberwithin{equation}{section}

\newcommand\eea{\end{eqnarray}}
\newcommand\bea{\begin{eqnarray}}

\def\beq{\begin{equation}}
\def\eeq{\end{equation}}

\newcommand{\be}{\begin{equation}}
\newcommand{\ee}{\end{equation}}
\newcommand{\ba}{\begin{align}}
\newcommand{\ea}{\end{align}}
\newcommand{\bg}{\begin{gather}}
\newcommand{\eg}{\end{gather}}
\newcommand{\bseq}{\begin{subequations}}
\newcommand{\eseq}{\end{subequations}}

\newcommand{\Det}{{\rm Det}}

\renewcommand{\t}{\tilde}

\newcommand{\coment}[1]{}

\begin{document}
\setcounter{page}0
\def\ppnumber{\vbox{\baselineskip14pt
}}
\def\ppdate{
} \date{}

\author{Germ\'an Mato$^{1,2}$, Facundo Rigatuso$^{1,2}$, Gonzalo Torroba$^{1, 2}$\\
[7mm] \\
{\normalsize \it $^1$ Instituto Balseiro, UNCuyo and CNEA}\\
{\normalsize \it $^2$Centro At\'omico Bariloche and CONICET}\\
{\normalsize \it S.C. de Bariloche, R\'io Negro, R8402AGP, Argentina}\\
}

\bigskip
\title{\bf  Dynamics of random recurrent neural networks: colored noise and sublinear activations
\vskip 0.5cm}
\maketitle

\begin{abstract}
We study the dynamics of recurrent neural networks with random couplings, sublinear odd activation functions, and temporally correlated noise. Using large $N$ methods, we determine stationary correlations and distinguish the loss of local linear stability from the onset of chaos, characterized by a positive largest Lyapunov exponent. These thresholds delimit an intermediate regime in which local instability coexists with asymptotic stability under a common noise history. Their separation depends on both the noise correlation time and the asymptotic growth of the activation function.  We show that internal noise with any finite correlation time cuts off the divergence of the network's intrinsic correlation time $\tau_\textrm{corr}$ at the edge of chaos; it exhibits a critical scaling $\tau_\textrm{corr,max} \sim 1/\sqrt{D_0}$, where $D_0$ is the integrated noise strength. Static random input instead produces coincident thresholds and a divergent correlation time. We also calculate the frequency-dependent participation dimension from a four-point correlation function. Its zero-frequency value is inversely proportional to the fourth power of the network correlation time, predicting a smooth minimum at the weak-noise chaos threshold. Numerical simulations support the large-network predictions and reveal appreciable finite-size effects.

\end{abstract}
\bigskip


\newpage
\tableofcontents

\vskip 1cm

\section{Introduction}\label{sec:intro}

Recurrent neural networks (RNNs) play a key role for various computational tasks in biological and artificial systems \cite{lipton2015criticalreviewrecurrentneural, 10.1162neco_a_01199}. An important goal is to understand their dynamics, given that the dynamical state of the system strongly influences its computational capabilities \cite{RevModPhys.78.1213}. 
The seminal work of Sompolinsky, Crisanti and Sommers \cite{PhysRevLett.61.259} established that large recurrent networks with random, asymmetric couplings and nonlinear activation can undergo a transition to chaos.  Crucially, networks that operate near the transition to chaos are found to perform optimally \cite{10.1162/089976604323057443, sussillo2009generating, PhysRevE.84.051908}. The dynamical mean-field theory of \cite{PhysRevLett.61.259}  characterized the activity correlations and largest Lyapunov exponent. Subsequent path-integral work provided a systematic derivation of the mean-field description and its stability properties \cite{Crisanti_2018}. These results made random recurrent networks a tractable setting for studying how microscopic interactions generate complex temporal activity.

Fluctuating inputs add important modifications to this problem. Correlation theory for stochastic neural networks showed how network interactions shape temporal and cross-neuron correlations \cite{ginzburg1994theory}. Additive noise was shown early to suppress chaos in random neural
networks \cite{PhysRevLett.69.3717}. Deterministic periodic input can also suppress intrinsic chaos, with
an effectiveness that depends nonmonotonically on the driving
frequency \cite{PhysRevE.82.011903}. For driven random networks,  \cite{Schuecker_2018} found that time-dependent internal white noise can suppress chaos and shift its onset to stronger recurrent coupling. Their analysis also revealed the existence of an intermediate regime between the onset of local instability and the transition to chaos. Thus, the response to a small perturbation and the long-term growth of differences between trajectories describe distinct aspects of the dynamics. The intermediate regime is locally unstable but nonchaotic. These two transitions also provide a useful framework for interpreting changes in population covariance and effective recurrent coupling \cite{shen2025covariance}.
Related work on spiking circuits identified mechanisms for asynchronous chaotic rate fluctuations with distinct characteristic time scales \cite{harish2015asynchronous}.

In this work we study two aspects of RNNs that can qualitatively affect the dynamics: the temporal structure of disordered input and the activation function. First, prior works have mostly focused on the case of internal white noise, which corresponds to a vanishingly small correlation time. However, realistic internal fluctuations arising from neural and synaptic processes can persist over a finite interval. We will model this with an  Ornstein--Uhlenbeck (OU) input that introduces a tunable correlation time $\tau_n$. Varying $\tau_n$ allows us to ask how internal colored noise affects network correlations, local instability and the onset of chaos.
Second, the activation function affects the dynamics in key ways. Many classical analyses use a bounded, saturating response, but other response shapes may be relevant over the range of activity sampled by the network. For example, type-I excitability is associated with a square-root dependence of firing rate on input near onset \cite{ermentrout1996type,gutkin1998dynamics,izhikevich2007dynamical}. Near critical behavior generated by linear activation functions has been found to be relevant for describing brain activity \cite{dahmen2020strong, dahmen2019second}. Previous work has shown that nonlinear response properties influence effective recurrent coupling and population correlations \cite{shen2025covariance, Mato:2025yof}. In this work we use a smooth family of activation functions that spans saturating and sublinear (nonsaturating) responses and approaches a linear response as a limiting case.

Our goal is to determine, both theoretically and via numerical simulations,
the dynamics of RNNs. In particular, we seek to understand how the colored
noise and the form of the activation function affect the unstable dynamics
and the development of chaos. We derive the large-network correlation
equations, and a matched short-/long-time solution that reproduces the
simulated correlators without measured input. We then examine two dynamical
thresholds: local instability, from the local stability criterion, and
chaos, from the sign of the largest Lyapunov exponent, which we obtain from
a Schr\"odinger problem whose potential is built from the correlator of
the derivative of the activation function. The two thresholds are distinct, with an intermediate regime that is
locally unstable but not chaotic, and simulations map their separation as a
function of the activation exponent and the noise correlation time.
Near the edge of chaos, internal noise of any finite correlation time cuts
off the divergence of the intrinsic correlation time with a universal
exponent, $\tau_\textrm{corr,max}\sim D_0^{-1/2}$, valid in the limit $D_0 \to 0$ of
integrated noise strength. For weak noise the maximum sits at the chaos
threshold. The exponent is the same for white and colored noise. We also investigate
whether these dynamical regimes differ in the geometric organization of
population activity, as characterized by the participation dimension
\cite{gao2017theory}. Its evaluation requires a four-point correlation
function, which captures the contribution of correlations between neurons
to the effective dimensionality of the activity. We find that the
zero-frequency participation dimension is controlled by the same intrinsic
time scale, $D_{PR}(0)/N=(\tau/\tau_\textrm{corr})^4$, so that it has a
minimum scaling as $D_0^2$ in the weak disorder limit. The simulations also allow us
to determine the relevance of finite-size effects in the different regimes.

The paper is organized as follows. Section~\ref{sec:model} introduces the random recurrent network, the colored-noise process, the family of activation functions, and the numerical methods. Section~\ref{sec:largeN} develops the large-$N$ mean-field theory, analyzes the temporal correlations, and derives criteria for local linear instability and chaos.
Section \ref{sec:weak-disorder} studies the dynamics in the weak disorder limit, obtaining explicit results for the instability and chaos thresholds and for the network correlation time. Section~\ref{sec:comparison} compares the theoretical predictions with simulations and presents phase diagrams showing how both transition thresholds depend on the noise correlation time and activation function. Section~\ref{sec:dpr} examines four-point correlations and the participation dimension of the network activity. Section~\ref{sec:discussion} contains a discussion of the results and conclusions. The appendices provide details of the collective-field and replica derivations and discuss finite-size effects.

\section{The model}\label{sec:model}

\subsection{Dynamics, disorder and internal noise}\label{subsec:dynamics}

We consider a network of $N$ recurrently coupled units with continuous time and a firing-rate dynamics. Denoting by $\phi_i(t)$ the total input current to unit $i$, and by $f(\phi)$ a single-neuron activation function, the network obeys the Langevin equation
\be\label{eq:eommodel}
\tau \frac{d\phi_i(t)}{dt} = -\phi_i(t) + \sum_{j=1}^N W_{ij}\, f\big(\phi_j(t)\big) + \xi_i(t)\,,\qquad i=1,\ldots, N\,,
\ee
where $\tau$ is the single-neuron integration time constant. This is a standard model used throughout the random-network literature; in the linear case it reduces to a multivariate Ornstein-Uhlenbeck process. See e.g. \cite{ginzburg1994theory, dahmen2020strong, dahmen2019second, PhysRevLett.61.259} for related rate-network dynamics and correlation theories. The model contains two independent sources of randomness, which play different roles: the \emph{$W$-disorder} and the \emph{internal noise}.

The recurrent couplings $W_{ij}$ are quenched (time-independent) random variables, drawn once per network realization from a Gaussian ensemble with
\be\label{eq:Wstats}
\langle W_{ij} \rangle = 0\,,\qquad \langle W_{ij} W_{kl} \rangle = \frac{\lambda^2}{N}\, \delta_{ik}\delta_{jl}\,.
\ee
The parameter $\lambda$ controls the strength of the recurrent coupling and is the main control parameter of the theory: $\lambda \to 0$ is the decoupled limit, while $\lambda\gg 1$ is the strong-coupling regime. Because $W$ does not evolve in time, its effect on the dynamics is entirely captured, at large $N$, by disorder averages $\langle \cdots \rangle_W$; this is the noise source responsible for generating an effective recurrent interaction in the single-neuron dynamics.

On the other hand, each neuron receives a time-dependent noise $\xi_i(t)$, uncorrelated across neurons,
\be
\langle \xi_i(t) \rangle = 0\,,\qquad \langle \xi_i(t)\,\xi_j(t')\rangle = \delta_{ij}\, G_\xi(t-t')\,.
\ee
This is a dynamical (annealed) noise, and unlike most previous literature on random networks, which specializes to white noise, we keep $G_\xi(t-t')$ general. This is what we will call \textit{colored noise}. A convenient choice, used throughout this work, is an Ornstein-Uhlenbeck noise,
\be\label{eq:colorednoise}
\frac{d\xi_i(t)}{dt} = -\frac{\xi_i(t)}{\tau_n} + \sigma_n\, \eta_i(t)\,,\qquad \langle \eta_i(t)\eta_j(t')\rangle = \delta_{ij}\,\delta(t-t')\,,
\ee
whose stationary two-point function is
\be\label{eq:colored}
G_\xi(t) = D_\xi\, e^{-|t|/\tau_n}\,,\qquad D_\xi \equiv \frac{\sigma_n^2 \tau_n}{2}\,.
\ee
We will denote
\be\label{eq:D0def}
D_0 \equiv \frac{1}{2} \int_{-\infty}^\infty dt\,G_\xi(t)=D_\xi \tau_n\,.
\ee
Two limits of \eqref{eq:colored} recover previously studied cases: the white-noise limit is $\tau_n\to 0$, $D_\xi \to \infty$ with $D_0$ fixed, and reduces to the case considered in \cite{Schuecker_2018,shen2025covariance}. The quenched limit is $\tau_n\to\infty$ at fixed $D_\xi$, in which each neuron sees an effectively static offset, $G_\xi(t)\to D_\xi$; this was studied in \cite{Mato:2025yof}. Away from these two limits, $\tau_n$ defines a new time scale of the problem, to be compared with $\tau$ and with the intrinsic correlation time generated self-consistently by the recurrent dynamics; one of our main goals below is to understand how this comparison controls the correlation structure and the onset of chaos. For a discussion of colored noise in the context of cortical neurons, and previous references, see \cite{10.1162/089976602320264015}

\subsection{Activation function}\label{subsec:activation}

We take $f(\phi)$ to be odd, $f(-\phi)=-f(\phi)$, sublinear at large $|\phi|$, and smooth. Building on our previous work \cite{Mato:2025yof}, we use a Pad\'e-type form
\be\label{eq:padesmoothmodel}
f(\phi)= \frac{\phi}{\left(1+\beta^2 \phi^2\right)^{\frac{1-p}{2}}}\,,\qquad 0\le p \le 1\,,\;\beta>0\,,
\ee
which has the asymptotic behavior
\begin{align}
f(\phi)&\approx \phi \;,\;\textrm{for}\; |\phi|\ll 1/\beta \nonumber\\
f(\phi) &\approx \beta^{p-1}\,\mathrm{sgn}(\phi)\,|\phi|^{p}\;,\;\textrm{ for}\; |\phi|\gg 1/\beta\,,
\end{align}
and is $C^\infty(\mathbb R)$ everywhere. This smoothness will be convenient for perturbative expansions and for the numerical evaluation of $f$ and its derivatives. The gain parameter $\beta$ sets the scale at which the nonlinearity turns on.

When $p \to 0$ this gives a saturating activation function, which we expect will have similar properties than other saturating functions such as $\tanh(x)$. In the limit $p \to 1$, we obtain the linear case. In between both limits, we obtain a nonlinear and nonsaturating regime. In realistic situations the activation function will eventually saturate; but we will assume that the neural activation function is well approximated by  (\ref{eq:padesmoothmodel}) over a range of parameters.  As a potential application, $p=1/2$ corresponds to neural dynamics that undergo a type I  bifurcation that leads to spike generation \cite{ermentrout1996type,gutkin1998dynamics,izhikevich2007dynamical}. 
One of the goals of this work is to explore this new parameter space. In particular, we will find an interesting interplay between $p$ and the dynamical evolution of the network.

\subsection{Numerical experiments}\label{subsec:numerics}

In this work we will use both numerical simulations and theoretical calculations to understand the dynamics of nonlinear RNNs. Before proceeding to a more detailed theoretical analysis, let us briefly discuss how \eqref{eq:eommodel} is numerically solved.

Eq. \eqref{eq:eommodel} together with the Ornstein-Uhlenbeck noise \eqref{eq:colorednoise} is integrated with a simple Euler-Maruyama scheme at fixed step $\delta t$,
\begin{align}
\xi_i(t+\delta t) &= \left(1-\frac{\delta t}{\tau_n}\right)\xi_i(t) + \sqrt{\delta t}\,\sigma_n\,\eta_i(t)\,,\qquad \eta_i(t)\sim\mathcal N(0,1)\ \text{i.i.d.}\,,\\
\phi_i(t+\delta t) &= \phi_i(t) + \frac{\delta t}{\tau}\Big[-\phi_i(t) + \sum_j W_{ij} f\big(\phi_j(t)\big) + \xi_i(t)\Big]\,.
\end{align}
The first line is the Ornstein-Uhlenbeck process discretized at step $\delta t$ up to $O(\delta t)$, with noise amplitude $\sigma_n=\sqrt{2D_\xi/\tau_n}$ fixed by \eqref{eq:colored}; we verify this normalization directly, by simulating a large population of independent OU processes with no recurrent coupling and measuring $\langle \xi_i(t)\xi_i(0)\rangle$, which reproduces $D_\xi e^{-|t|/\tau_n}$ to high accuracy. The couplings $W_{ij}$ are drawn once per realization from \eqref{eq:Wstats}. The activation function is \eqref{eq:padesmoothmodel} at fixed gain $\beta$; the exponent $p$ is treated as an external parameter of the numerical experiment, and $\lambda$ is scanned over a grid at fixed $p$.

For each realization of $W$ (a ``$W$-seed'') the dynamics is run for a transient of $\delta t\, n_{\rm tran}$ time units to reach the stationary state, then for a further $\delta t\, n_{\rm max}$ time units during which the trajectory $\phi_i(t)$ is recorded. Repeating this for several independent noise realizations (``replicas'') at fixed $W$, and averaging the population autocorrelation over neurons and replicas, gives numerical estimates of
\be
C_\phi(t) = \frac{1}{N}\sum_i \langle \phi_i(t+t')\phi_i(t')\rangle\,,\qquad
C_f(t) = \frac{1}{N}\sum_i \langle f(\phi_i(t+t')) f(\phi_i(t'))\rangle\,,
\ee
computed via FFT-based autocorrelation (and similarly for other correlators that will be needed). We have checked that after these averages and the transient, these correlators are independent of $t'$, signaling a time-translation invariant state.

An important goal of this work will be to analyze the transition to chaos and construct a phase diagram distinguishing the stable, unstable, and chaotic regimes. The onset of the chaotic regime is defined by the largest Lyapunov exponent $\Lambda_L$ becoming positive. We will give more details of the theoretical calculations below in Sec.~\ref{subsec:chaos}. The largest Lyapunov exponent is estimated with the standard Benettin two-replica algorithm \cite{benettin1980lyapunov}:  two copies of the network are evolved under the same realization of $W$ and driven by the same noise history $\xi_i(t)$ (i.e.\ the same $\eta_i(t)$ at every step), starting from initial conditions separated by a small distance $d_0$; after each step the separation $d(t)=\|\phi^{(1)}(t)-\phi^{(2)}(t)\|$ is measured and the second trajectory is rescaled back onto the sphere of radius $d_0$ around the first, accumulating $\log(d/d_0)$. After discarding an initial transient, the running time-average of this quantity converges to $\Lambda_L$.

The scan is organized as a double loop over network size $N$ and coupling $\lambda$ at fixed $p$ (with $\beta$, $D_\xi$, $\tau_n$, $\tau$ fixed); for each $(N,\lambda)$ several independent $W$-realizations are drawn, and for each $W$ several noise replicas are run. This nested structure allows the disorder-vs-noise decomposition of the variance of any observable $X$ (e.g.\ $C_\phi(0)$ or $\Lambda_L$) into a between-$W$ (quenched) component and a within-$W$, between-replica (annealed) component, which is the numerical counterpart of the self-averaging property assumed in the large-$N$ derivation of Sec.~\ref{sec:largeN}. Additional diagnostics recorded for each $(N,\lambda)$ include the spread of the time-averaged activity $\bar\phi_i$ across neurons (a check that no neuron is atypical), and the power spectrum of $\phi_i(t)$ (used to detect any oscillatory instability).

\section{Theoretical analysis at large $N$}\label{sec:largeN}

In this section we derive the large-$N$ dynamics of \eqref{eq:eommodel} using the Martin-Siggia-Rose/De Dominicis-Janssen path integral; see e.g. \cite{Helias_2020} for a review.

\subsection{Path integral and the self-consistent equation}\label{subsec:selfconsistent}

 The random $W$ and $\xi$ dynamics is given in terms of appropriate Gaussian distributions; the neural network equations are imposed via a Lagrange multiplier $\t \phi$, and we assume an Ito-type discretization that leads to a trivial Jacobian determinant. The path integral representation in the time domain is then 
\begin{align}\label{eq:action1model}
Z&= \int DW D\xi D\phi D \t \phi \,e^{-\frac{N}{2\lambda^2} \sum_{ij} W_{ij}^2} e^{-\frac{1}{2} \int_{t,t'} \sum_i \xi_i(t) G_\xi^{-1}(t,t') \xi_i(t')} \nonumber\\
&\qquad \times\;e^{i \int_t \sum_i\t \phi_i(t) \left[ (\tau \partial_t+1) \phi_i(t)-\sum_j W_{ij} f(\phi_j(t))-\xi_i(t)\right]}\,.
\end{align}
Integrating over $W$, $\xi$ and $\t \phi$ we arrive at
\be
Z= \int D\phi\, [{\rm Det}(2\pi M^{-1})]^{N/2} e^{-\frac{1}{2} \int_{t,t'} \sum_i (\tau \partial_t+1) \phi_i(t) M^{-1}(t,t') (\tau \partial_{t'}+1) \phi_i(t')}
\ee
where
\be\label{eq:Mdefmodel}
M(t,t') = G_{\xi}(t-t')+ \frac{\lambda^2}{N} \sum_j f(\phi_j(t)) f(\phi_j(t'))\,.
\ee
Integrating by parts in the exponential, this can be written as
\be
Z= \int D\phi\, [{\rm Det}(2\pi M^{-1})]^{N/2} e^{-\frac{1}{2} \int_{t,t'} \sum_i  \phi_i(t) G^{-1}(t,t') \phi_i(t')}\,,
\ee
and a short calculation gives the following relation between the $\phi$ 2-point function $G(t,t')$ and $M(t,t')$
\be
(\tau \partial_t+1)(\tau \partial_{t'}+1) G(t,t') = G_{\xi}(t-t')+ \frac{\lambda^2}{N} \sum_j f(\phi_j(t)) f(\phi_j(t'))\,.
\ee

At large $N$, the expectation value of this relation gives the single-site self-consistent equation
\be\label{eq:self1model}
(\tau \partial_t+1)(\tau \partial_{t'}+1) G(t,t') = G_{\xi}(t-t')+\lambda^2 \langle f(\phi(t)) f(\phi(t')) \rangle_G\,.
\ee
The single site correlators are
\be
G(t,t')= \langle \phi(t) \phi(t') \rangle =\frac{1}{N}\sum_j\langle \phi_j(t) \phi_j(t') \rangle 
\ee
and
\begin{align}\label{eq:Cfdefmodel}
\langle f(\phi(t)) f(\phi(t')) \rangle_G&=\frac{1}{N} \sum_j \langle f(\phi_j(t)) f(\phi_j(t'))\rangle \nonumber\\
&= \int \frac{D\phi }{\sqrt{{\rm Det}(2\pi G)}}\,f(\phi(t)) f(\phi(t')) e^{-\frac{1}{2} \int_{t,t'}  \phi(t) G^{-1}(t,t') \phi(t')}\,.
\end{align}

In our numerical simulations of the neural network, we have checked that after a transient the network is described by a time-translation invariant configuration. Assuming this in the present theoretical derivation, we can look for a nontrivial self-consistent solution that depends only on $t-t'$.
We denote the time-translation invariant self-consistent two-point functions by
\begin{align}\label{eq:defCphiCfmodel}
C_\phi(t-t')&\equiv G(t,t')= \langle \phi(t)  \phi(t')\rangle\,, \,
C_f(t-t')\equiv \langle f\left(\phi(t)\right)  f\left(\phi(t')\right)\rangle_{C_\phi(t-t')}\,,
\end{align}
where from now on we will keep implicit the sum over sites and the $1/N$ normalization as in (\ref{eq:Cfdefmodel}). The self-consistent equation then becomes
\be\label{eq:selfrepeat}
(-\tau^2 \partial_t^2+1) C_\phi(t)-\lambda^2 C_f(t) = G_{\xi}(t)\,,
\ee
which should be solved together with the initial and final conditions
\be\label{eq:initialc}
\dot C_\phi(t=0)=0\;,\;\lim_{t \to \infty} C_\phi(t)=0\,.
\ee
(We take $G_\xi(t)$ even in $t$ and $f(\phi)$ odd in $\phi$). This is one of the main theoretical results.
It extends previous white-noise formulas \cite{Schuecker_2018, Helias_2020, grosvenor2022edgechaosquantumfield, shen2025covariance} to a general colored noise $G_\xi(t)$. Setting $G_\xi(t)=2 D_0\delta(t)$ recovers the white-noise self-consistent equation,\footnote{In this limit, $\dot C_\phi(t)$ becomes discontinuous at $t=0$.} and $G_\xi(t)=D_q$ recovers the static (quenched) noise.

Eq. \eqref{eq:selfrepeat} can equivalently be obtained, at the same order in $1/N$, from a collective-field representation of the disorder average. We present this alternative derivation in Appendix~\ref{app:collective}, since it also introduces the notation needed later for the replica calculation of the 4-point function.

For future use, let us give the explicit integral for $C_f(t)$. Writing
\be
z_1 = C_\phi(0)^{-1/2}\,\phi(t)\,,\qquad z_2 = C_\phi(0)^{-1/2}\,\phi(0)\,,\qquad u(t)=\frac{C_\phi(t)}{C_\phi(0)}\,,
\ee
so that $z_1,z_2$ are unit-variance Gaussians with correlation $u(t)$, their joint density is 
\be\label{eq:Pkernelmodel}
P(z_1,z_2) = \frac{1}{2\pi\sqrt{1-u(t)^2}}\exp\left[-\frac{z_1^2+z_2^2-2u(t)\,z_1z_2}{2(1-u(t)^2)}\right]\,.
\ee
In terms of this kernel, we have
\be\label{eq:CfPmodel}
C_f(t) = \int dz_1\,dz_2\; P(z_1,z_2)\,f\!\left(\sqrt{C_\phi(0)}\,z_1\right) f\!\left(\sqrt{C_\phi(0)}\,z_2\right)\,,
\ee
where both $z_1$ and $z_2$ take values in $\mathbb R$. If the correlator is taken at equal times, the expression becomes instead a single integral,
\be\label{eq:Cf0intmodel}
C_{f}(t=0)=\langle f(\phi(t)) f(\phi(t)) \rangle_{C_\phi(0)}=\int dz \frac{e^{-z^2/2}}{\sqrt{2\pi}}\left[f\left(\sqrt{C_\phi(0)}z \right)\right]^2\,.
\ee
Below in some formulas we will need to perform Taylor expansions and take variations with respect to $C_\phi(t)$. It will then be useful to emphasize that $C_f(t)$ is computed from two variables $\phi(t)$ and $\phi(0)$ with 2-point functions $C_\phi(t)$ and $C_\phi(0)$; we will denote this by $C_f\left( C_\phi(t), C_\phi(0)\right)$.

\subsection{Approximate explicit solution and stability criterion}\label{subsec:approxsolution}

Eq. \eqref{eq:selfrepeat} is a nonlinear integro-differential equation for $C_\phi(t)$ and cannot be solved in closed form. However, we will see that simple short and long time expansions are sufficient for capturing the main properties. We now describe how these arise.

Let us first study the short-time behavior of \eqref{eq:selfrepeat}, which as we shall now see, will also help us distinguish between the stable and unstable regime. We write 
\be
q\equiv C_\phi(0)\,,
\ee 
and recall that $\dot C_\phi(0)=0$.
By short time expansion, we mean a time window $t \in (0, t_*)$ where the following expansion is valid:
\be
C_\phi(t) \approx q + \hat C_\phi(t)\;,\; |\hat C_\phi| \ll q\,.
\ee

As we discuss below, the constant $q$ will be determined globally, by matching onto the long-time expansion and requiring $C_\phi(t \to \infty)=0$. Using Price's theorem\footnote{This follows by integrating by parts inside Gaussian integrals.}
\be
\frac{\partial}{\partial C_\phi(t)} C_f(C_\phi(t), C_\phi(0))= C_{f'}(C_\phi(t), C_\phi(0))
\ee
(where $f'(\phi)=\partial_\phi f(\phi)$)
we have
\be
C_f(t)\approx \langle f^2 \rangle_q+\langle (f')^2\rangle_q \hat C_\phi(t)\,.
\ee
The notation here is $\langle f^2 \rangle_q= \langle f(\phi(t)) f(\phi(t))\rangle_q$; we are suppressing the time dependence since in the time translation invariant state in which we are working, the expectation value is independent of $t$.
Then, introducing
\be
\Omega^2 \equiv \frac{\lambda^2 \langle (f')^2 \rangle_q-1}{\tau^2}\,,
\ee
the self-consistent equation gives the following equation at leading order in the fluctuation $\hat C_\phi$:
\be
- (\partial_t^2 +\Omega^2) \hat C_\phi(t) = \frac{G_\xi(t) + \lambda^2 \langle f^2\rangle_q -q}{\tau^2}\,.
\ee

We see that $\Omega^2$ controls the stability of linear fluctuations: for $\Omega^2<0$ the linear fluctuations that do not grow at infinity decrease exponentially fast, while for $\Omega^2>0$ we enter a regime where the homogeneous solutions oscillate. Once $q$ is determined self-consistently, the transition $\Omega^2=0$ happens at a specific value of the coupling, which we denote by $\lambda_{c,1}$:
\be\label{eq:lambdac1}
\lambda_{c,1}^2 \langle (f')^2 \rangle_q-1=0\,.
\ee

The short-time solution is determined as an
initial-value problem, with $\hat C_\phi(0)=\dot{\hat C}_\phi(0)=0$. It is useful to introduce the causal kernel $K(t)$, defined by 
\be
\ddot K+\Omega^2K=0\;,\;K(0)=0\;,\;\dot K(0)=1\,.
\ee
It is given by
\be
K(t)=\frac{\sin(\Omega t)}{\Omega}\quad(\Omega^2>0)\,,\qquad
K(t)=\frac{\sinh(|\Omega| t)}{|\Omega|}\quad(\Omega^2<0)\,.
\ee
The short-time solution then reads
\be\label{eq:Chatshort}
\hat C_\phi(t)=-\frac{1}{\tau^2}\int_0^t dt'\,K(t-t')\Big[G_\xi(t')+\lambda^2\langle f^2\rangle_q-q\Big]\,.
\ee

On the other hand, at long times the correlator $C_\phi(t)$ decays and it is convenient to expand $C_f(t)$ in powers of $C_\phi(t)/q$. To obtain this, we expand the kernel (\ref{eq:Pkernelmodel}) using Mehler's formula,
\be
P(z_1,z_2) =\frac{e^{-z_1^2/2}}{\sqrt{2\pi}}\frac{e^{-z_2^2/2}}{\sqrt{2\pi}} \sum_{k=0}^\infty\,\frac{u(t)^k}{k!} \textrm{He}_k(z_1)\textrm{He}_k(z_2)\,,
\ee
where $ \textrm{He}_k(z)$ are the probabilistic Hermite polynomials.
Introducing the Hermite coefficients of the activation function, 
\be
f_k(q)=\int dz\,\frac{e^{-z^2/2}}{\sqrt{2\pi}}\,\textrm{He}_k(z)\,f(q^{1/2}z)\,,
\ee 
we find
\be\label{eq:mehler}
C_f(t)= \sum_{k=0}^\infty\,\frac{f_k(q)^2}{k!} \left(\frac{C_\phi(t)}{q}\right)^k= \sum_{k=0}^\infty \frac{1}{k!}\,\langle f^{(k)}(\phi)\rangle^2_{q}\,C_\phi(t)^k\,.
\ee
The second equality is obtained using Rodrigues' formula and integration by parts (this requires $C^\infty$ smoothness of $f$). Since $f$ is odd, it receives contributions only from odd $k$. 

Keeping the first term, the long-time self-consistent equation reduces to
\be\label{eq:longtimes1}
-\ddot C_\phi(t) + \kappa^2 C_\phi(t)\approx \frac{1}{\tau^2} G_\xi(t)\,,
\ee
with
\be\label{eq:kappa}
\kappa(q)^2 \equiv \frac{1-\lambda^2 \langle f' \rangle^2_q}{\tau^2}\,.
\ee
The parameter $\kappa$ fixes the decay rate for the homogeneous solution at long times; using the Cauchy-Schwarz inequality gives $\kappa^2 \ge 0$, and $\kappa$ can vanish only if $f(\phi)=\phi$. See \cite{shen2025covariance} for a proof of this point. We will define the \textit{correlation time} of the RNN as
 \be\label{eq:taucorr}
 \tau_\textrm{corr}= \frac{1}{\kappa} = \frac{\tau}{\sqrt{1- \lambda^2 \langle f' \rangle_q^2}}\,.
 \ee
 This is an important observable that characterizes the dynamics of the network, and we will study it in more detail in the following sections. The solution to (\ref{eq:longtimes1}) is
\be\label{eq:longtimes2}
C_\phi(t) \approx \frac{D_\xi}{\tau^2} \frac{ \tau_n^2}{\tau_n^2 \kappa^2-1}  e^{-|t|/\tau_n}+B e^{-\kappa |t|}\,,
\ee
with $B$ an undetermined constant. Higher order contributions can be incorporated systematically using (\ref{eq:mehler}).

The final step is to match both regimes.\footnote{We have given explicit formulas including the first terms in each regime, but in numerical implementations we also include the $C_\phi^3$ term in the Mehler expansion.}
The short-time solution \eqref{eq:Chatshort} depends on the unknown equal-time value $q=C_\phi(0)$, and the long-time solution \eqref{eq:longtimes2} depends on $q$ (through $\kappa=\kappa(q)$) and on $B$. We determine both unknowns by matching the two solutions at a time $t^*$ in their common range of validity, requiring continuity of the correlator and of its first derivative. The matching time is constrained by requiring that both solutions be accurate on their side of $t^*$; within this range, $t^*$ is chosen to minimize the difference of second derivatives. We will find in Sec. \ref{sec:comparison} a very good agreement between this matching procedure and the simulation results.

\subsection{Chaotic dynamics}\label{subsec:chaos}

We have already characterized theoretically the unstable regime in terms of the critical coupling $\lambda_{c,1}$ defined in (\ref{eq:lambdac1}), and now we will focus on the transition to chaos. For this, we need to analyze the sensitivity of the dynamics \eqref{eq:eommodel} to initial conditions. Following \cite{PhysRevLett.61.259,Schuecker_2018,Crisanti_2018}, we consider two identical copies of the network with slightly different initial conditions, $\phi^{1,2}_i(t)$. They evolve under the same realization of $W$ and are driven by the same noise history $\xi_i(t)$. The largest Lyapunov exponent is
\be\label{eq:LLsim}
\Lambda_L= \lim_{t \to \infty} \lim_{|| \phi^1(0) - \phi^2(0)||\to 0}\,\frac{1}{2t}\,\log \frac{|| \phi^1(t) - \phi^2(t)||^2}{|| \phi^1(0) - \phi^2(0)||^2}\,.
\ee
The dynamics is chaotic if $\Lambda_L>0$.

The double-copy path integral gives a two-replica version of \eqref{eq:self1model} \cite{Schuecker_2018},
\be\label{eq:self2model}
(\tau \partial_t+1)(\tau \partial_{t'}+1) C_\phi^{\alpha \beta}(t,t') = G_{\xi}(t-t') +\lambda^2 C_f^{\alpha \beta}(t,t')\,,
\ee
where $\alpha, \beta=1,\,2$ are the replica indices, which distinguish the two different initial conditions.
Calling $C_\phi=C_\phi^{11}=C_\phi^{22}$ and expanding 
\be
C_\phi^{12}=C_\phi+\epsilon^2 C^{(1)}\;,\;\epsilon=\|\phi^1(0)-\phi^2(0)\|\to0
\ee 
we find that $C_\phi(t)$ obeys the ordinary self-consistent equation \eqref{eq:selfrepeat}, while the perturbation obeys
\be\label{eq:self4model}
(\tau \partial_t+1)(\tau \partial_{t'}+1) C^{(1)}(t,t')=  \lambda^2 C_{f'}(t,t') C^{(1)}(t,t')\,,
\ee
with $C_{f'}(t,t')=\langle f'(\phi(t))f'(\phi(t'))\rangle$ obtained via Price's theorem. 

As shown in \cite{Schuecker_2018}, we can write
\be
C^{(1)}(t,t')\sim e^{\Lambda_L t^+}\Psi(t^-)\;,\;t^\pm=t\pm t'\,,
\ee 
arriving at an effective Schr\"odinger problem
\be\label{eq:Schr2model}
\left(-\tau^2 \partial_t^2 + 1-\lambda^2 C_{f'}(t)\right)\Psi(t) = E\, \Psi(t)\,,\,E=1-(1+\Lambda_L\tau)^2\,.
\ee
This is a one-dimensional, time-independent Schr\"odinger equation in which $t$ plays the role of a spatial coordinate, and
\be\label{eq:Schr2model2}
W(t)= 1-\lambda^2 C_{f'}(t)
\ee
is the Schr\"odinger potential. In this representation, the chaotic regime corresponds to the existence of a bound state with $E<0$. This generalizes the analyses of \cite{PhysRevLett.61.259} (no internal noise) and \cite{Schuecker_2018} (white noise) to a general colored noise, which enters only through the self-consistently determined $C_{f'}(t)$. We define $\lambda_{c,2}$ as the value of the coupling at which the RNN transitions to the chaotic phase,
\be\label{eq:lambdac2def}
E(\lambda_{c,2})=0\,.
\ee

Taking into account that
\be
\frac{1}{\int dt\,|\Psi(t)|^2}\,\int dt\, \Psi^*(t)\left(-\tau^2\partial_t^2+W(t)\right) \Psi(t)=E \ge \min_t W(t)\,,
\ee 
a necessary condition for chaos, $E<0$, is 
\be
\min_t W(t)\le E<0\,.
\ee 
Since the two-point function $C_{f'}(t)$ decays with $t$, this minimum is at $t=0$. The necessary condition for chaos then becomes
\be\label{eq:necessary1model}
1-\lambda^2 C_{f'}(0)\le 0\,.
\ee
This coincides with the instability condition $\lambda> \lambda_{c,1}$ given in (\ref{eq:lambdac1}). Sufficient conditions can be derived by using trial wavefunctions because the energy ground state minimizes the energy. For instance, using $\psi_{tr}(t)=e^{-\kappa_{tr}|t|}$, the sufficient condition becomes
\be\label{eq:sufficientmodel}
\kappa_{tr}^2\tau^2+1-2\kappa_{tr}\lambda^2\int_0^\infty dt\, e^{-2\kappa_{tr} t}\, C_{f'}(t) \le 0\quad\text{for some }\kappa_{tr}>0\,.
\ee
From this analysis, we see that in general
\be
\lambda_{c,1} \le \lambda_{c,2}\,.
\ee
We will see in Sec. \ref{sec:weak-disorder} that the inequality is saturated only if $\dot G_\xi(t)=0$, or for white noise only for a linear activation function.

These necessary and sufficient conditions allow to distinguish the chaotic regime in semi-analytic but approximate ways. In what follows, we will proceed in two steps. First, in Sec. \ref{sec:weak-disorder} we will determine the main dynamical properties of RNN within a weak disorder expansion. This will provide explicit expressions for various quantities of interest, as well as establishing a direct connection between the chaos transition and soliton theory. Next, in Sec. \ref{sec:comparison} we will focus on solving the Schr\"odinger problem (\ref{eq:Schr2model}) numerically. Once $C_{f'}(t)$ has been computed, this is not hard to do by discretizing the time coordinate into $N_t$ steps, and diagonalizing the discrete Hamiltonian. We will find an excellent agreement between this theoretical result and the Lyapunov exponent (\ref{eq:LLsim}) from numerical simulations of the RNN.

\section{Dynamics of RNNs in the weak-disorder expansion}
\label{sec:weak-disorder}

In this section we study the self-consistent problem analytically in the
regime where the equal-time variance $q=C_\phi(0)$ is small, which is the
relevant regime close to the instability and chaos thresholds when the internal noise is weak.  We
treat in turn the noiseless case, white noise and quenched (static) noise. In particular, we will characterize the transition to chaos in the limit of vanishing disorder in terms of a critical exponent for the correlation time of the network, and we will compute analytically the value of this exponent.
In the last part of the section we will establish an explicit map between these results and a soliton solution and its fluctuations.

\subsection{Particle picture}

As discussed in \cite{PhysRevLett.61.259}, it is useful to view the self-consistent equation as motion of a particle in a potential and use energy conservation arguments. We now extend this to general colored noise. For this purpose, let us rewrite the self-consistent equation as

\begin{equation}\label{eq:wd-particle}
\tau^2\,\ddot C_\phi(t) =- \partial_{C_\phi(t)} V(C_\phi (t), q) - G_\xi(t),\qquad
V(C, q)=-\tfrac12 C^2+\lambda^2 C_F(C,\,q),
\end{equation}

where $F(x)=\int_0^x f$ and $C_F(C, q)=\langle F(x_1)F(x_2)\rangle$ for a
 Gaussian pair with $\langle x_i^2\rangle=q$ and
$\langle x_1x_2\rangle=C$.  In the mechanical analogy, this corresponds to a particle with position $C_\phi(t)$ moving in a potential $V$, and where the noise $G_\xi(t)$ provides a time-dependent driving force. The initial condition $\dot C_\phi(0)=0$ means the particle starts at $t=0$ at rest at some $q=C_\phi(0)$, and reaches $C_\phi=\dot C_\phi=0$ as $t \to \infty$ for a decaying $G_\xi(t)$ .

In the absence of noise, multiplying (\ref{eq:wd-particle}) by $\dot C_\phi(t)$ and integrating over time leads to a conserved energy,
\be
 \frac{\tau^2}{2} \dot C_\phi(t)^2+ V\left(C_\phi(t), q \right)=  \frac{\tau^2}{2} \dot C_\phi(t_0)^2+ V\left(C_\phi(t_0), q \right)
\ee
Using this, $q=C_\phi(0)$ is obtained by choosing $t=0$ and $t_0 \to \infty$:
\be\label{eq:q-nonoise}
V(q,q)= V(0,q)\,.
\ee
On the other hand, for white noise, $G_\xi=2D_0\delta(t)$ produces a kink,
$\dot C_\phi(0^\pm)=\mp D_0/\tau^2$, and energy conservation gives \cite{Schuecker_2018}
\begin{equation}\label{eq:wd-white-balance}
-V(q, q)+V(0, q)=\frac{D_0^2}{2\tau^2}.
\end{equation}

More generally, for colored noise the energy difference becomes
\be
\left[\frac{\tau^2}{2} \dot C_\phi(t')^2+V\left( C_\phi(t'), q\right) \right]_{t'=t_0}^{t'=t}= -\int_{t_0}^t dt' \,G_\xi(t') \dot C_\phi(t')\,.
\ee
This integral of the self-consistent equation in general does not offer additional simplifications. However, it becomes useful in the quenched limit, $G_\xi(t) \to D_q>0$ (constant), because the noise term can be integrated explicitly,
\be
 \frac{\tau^2}{2} \dot C_\phi(t)^2+ V\left(C_\phi(t), q \right)+ D_q C_\phi(t)=  \frac{\tau^2}{2} \dot C_\phi(t_0)^2+ V\left(C_\phi(t_0), q \right)+ D_q C_\phi(t_0)\,.
\ee
In this case, $C_\phi(t) \to C_\infty \neq 0$ for $t \to \infty$ due to the nondecaying noise. This value is fixed by evaluating the self-consistent equation as $t \to \infty$:
\begin{equation}\label{eq:Cinfty}
C_\infty=\lambda^2C_f(C_\infty, q)+D_q,\qquad C_\infty\ge D_q>0\,.
\end{equation}
The energy formula for $t_0=0$ and $t \to \infty$ then gives
\begin{equation}\label{eq:Energyquenched}
-V(q, q)+V(C_\infty, q)=D_q\,(q-C_\infty) .
\end{equation}

We will obtain analytic results on the dynamics of RNNs by combining a weak disorder expansion together with expansions of $V(C_\phi, q)$ for $C_\phi \to 0$ or $C_\phi \to q$. In order to give these expansions, we will use Price's theorem, 
 $\partial_{C_\phi} C_g = C_{g'}$,\footnote{The notation is as before, $C_g(t) = \langle g(\phi(t)) g(\phi(0)) \rangle$, $C_\phi(t)= \langle \phi(t) \phi(0) \rangle$, and $g'(t)= \partial_{\phi(t)} g(\phi(t))$.} and
Mehler's formula,
\begin{equation}
C_g(C_\phi, q)=\sum_{k\ge0}\frac{\langle g^{(k)}\rangle_q^2}{k!} C_\phi^k\,.
\label{eq:wd-mehler}
\end{equation}
 
For the expansion around $C_\phi=0$,  using Price's and Mehler's formula, and $f'(0)=1$, 
we obtain
\begin{equation}\label{eq:wd-V-u0}
V(C_\phi, q)=\lambda^2\langle F\rangle_q^2-\frac{1}{2}\,\tau^2\kappa^2 C_\phi^2
+\frac{\lambda^2\langle f^{(3)}\rangle_q^2}{24}\,C_\phi^4 +\dots,
\end{equation}
with $\tau^2\kappa^2=1-\lambda^2\langle f'\rangle_q^2$ as before.  Note that the constant $V(0, q)=\lambda^2\langle F\rangle_q^2$ does not vanish in general. 
For the expansion around $C_\phi=q$ we obtain
\be\label{eq:wd-V-u1}
V(C_\phi, q)=V(q, q)+\big[q-\lambda^2\langle f^2\rangle_q\big] (q-C_\phi)
-\frac{1}{2}\big[1-\lambda^2\langle f'^2\rangle_q\big] (q-C_\phi)^2+\dots
\ee
We note that the slope at $C_\phi=q$ vanishes when $q=\lambda^2\langle f^2\rangle_q$ and the
curvature vanishes when $\lambda^2\langle f'^2\rangle_q=1$, i.e.\ when
$\Omega^2=0$.  

\subsection{Noiseless case}\label{subsec:noiseless}
 
With $G_\xi=0$  energy
conservation gives $V(q, q)=V(0, q)$.  For $\lambda< 1/f'(0)=1$, the only solution is $q=0$. To see this, we use Price's theorem to expand at small $q$:
\begin{align}\label{eq:smallqV}
V(q,q)-V(0,q)&= -\frac{1}{2}q^2+\lambda^2 \left( \langle F^2 \rangle_q-\langle F \rangle_q^2\right)\\
&=- \frac{1}{2}(1-\lambda^2)q^2+ \frac{\lambda^2}{2} f^{(3)}(0) q^3  +\lambda^2 \left[\frac{1}{6}(f^{(3)}(0))^2+ \frac{1}{8} f^{(5)}(0) \right] q^4+ \ldots\,. \nonumber
\end{align}
We used that $f$ is odd and $C^\infty$; in particular, all even derivatives vanish at the origin.

For $\lambda<1$, $q=0$ is the unique stable solution \cite{PhysRevLett.61.259} (recall that $f^{(3)}(0)<0$). But we see that a nontrivial $q$ develops if $\lambda>1$; in particular, for $\lambda$ slightly above $1$,
\begin{equation}
q_0\simeq\frac{\lambda^2-1 }{\lambda^2 |f^{(3)}(0)|}\,.
\end{equation}
Since $\langle f'^2\rangle_q \to 1$ as $q \to 0$, the appearance of the nontrivial root $q_0$ coincides with the onset of linear instability, $\lambda^2 \langle f'^2 \rangle -1 =0$. The linear instability is present for all $\lambda >1$ and hence $\lambda_{c,1}=1$.\footnote{This can be shown by writing $\lambda^2$ as a function of $q$ from $V(q,q)=V(0,q)$ and then realizing that for such a $\lambda$, $\lambda^2  \langle f'^2 \rangle> 1$ in terms of the Cauchy-Schwarz inequality or with a Hermite basis expansion.}

Differentiating Eq.~(\ref{eq:wd-particle}) gives
\begin{equation}
\big[-\tau^2\partial_t^2+1-\lambda^2C_{f'}(t)\big]\dot C_\phi=0 ,
\label{eq:wd-zero-mode}
\end{equation}
so $\dot C_\phi$ is an $E=0$ eigenstate of the Schr\"odinger problem.  In
the noiseless case it is odd with a node at $t=0$, hence the ground state
has $E_0<0$ and the network is chaotic.  This reproduces the result of \cite{PhysRevLett.61.259},
\be\label{eq:SCSlambda}
\lambda_{c,1}=\lambda_{c,2}=1\,.
\ee
 
 The phase transition can be further characterized in terms of the correlation time (\ref{eq:taucorr}). Writing this in terms of $q_0$ and expanding for $\lambda \sim 1$, we obtain
 \be\label{eq:taucorrnoiseless}
 \tau_\textrm{corr}= \frac{\tau}{\sqrt{1- \lambda^2 \langle f' \rangle_q^2}} \sim \frac{\tau}{|\lambda-1|}\,.
 \ee
 At the transition between the stable and chaotic phases the correlation time diverges in the noiseless case. We will see that a time-dependent input noise cuts off this divergence.
 
\subsection{White and colored noise}\label{subsec:whitenoise}

We first derive the weak-noise behavior for white noise, where the energy balance is exact, and then argue that the main conclusions carry over to colored noise.

For white noise, $G_\xi=2D_0\delta(t)$, energy conservation gives (\ref{eq:wd-white-balance}), so $q$ is always nonvanishing, even at weak coupling. The expansion in small $q$ used below is controlled by $|f^{(3)}(0)|\,q\ll 1$; we will see that this requires $D_0\ll \tau/|f^{(3)}(0)|$. Since $f'(0)=1$, $|f^{(3)}(0)|$ measures the curvature (saturation) of the activation function; for the Pad\'e form, $|f^{(3)}(0)|=3(1-p)\beta^2$. If $D_0$ is not small in this sense, nonlinear effects of the activation function are important and $q$ can only be found numerically; we explore this general case in Sec.~\ref{sec:comparison}. Here we take the weak-noise limit and determine the resulting dynamical transitions.

For weak noise and small $\lambda$ we can use the small-$q$ expansion (\ref{eq:smallqV}), and so we have to solve
\be\label{eq:wd-white-balance2}
\frac{1}{2}(1-\lambda^2)q^2- \frac{\lambda^2}{2} f^{(3)}(0) q^3  -\lambda^2 \left[\frac{1}{6}(f^{(3)}(0))^2+ \frac{1}{8} f^{(5)}(0) \right] q^4+ \ldots=\frac{D_0^2}{2\tau^2}\,.
\ee
When $\lambda \ll 1$ the solution is dominated by the linear dynamics,
\be
q\simeq\frac{D_0}{\tau \sqrt{1-\lambda^2}}\,.
\ee
Near $\lambda=1$, on the other hand, it is dominated by the cubic term,
\be
q \simeq \frac{D_0^{2/3}}{\tau^{2/3} \lambda^{2/3} |f^{(3)}(0)|^{1/3}}\,.
\ee
The scaling with $D_0$ changes but, in contrast to the noiseless case, there is no transition at $\lambda=1$. In both regimes the condition $|f^{(3)}(0)|\,q\ll1$ is equivalent to
\be\label{eq:weak-noise}
D_0 \ll \frac{\tau}{|f^{(3)}(0)|}\,,
\ee
which is the quantitative statement of the weak-noise limit. We will see below that the same condition ensures that $q$ is small at the instability and chaos thresholds. For activation functions with order-one curvature this is simply $D_0\ll\tau$.

Next we consider the linear instability (\ref{eq:lambdac1}). At weak noise it occurs near $\lambda=1$ and for small $q$. Expanding (\ref{eq:lambdac1}) at small $q$,
\be\label{eq:lambdac11}
\lambda_{c,1}^2=\frac{1}{ \langle (f')^2 \rangle_q}= 1 - f^{(3)}(0) q+ \frac{1}{4} \left[(f^{(3)}(0))^2-f^{(5)}(0) \right]q^2 + \ldots
\ee
Then
\be
V(q,q)-V(0,q) \Big|_{\lambda=\lambda_{c,1}}=- \frac{5}{24} \left( f^{(3)}(0) \right)^2 q^4+ O(q^5)\,,
\ee
and so the self-consistent solution occurs at
\be
q(\lambda_{c,1})\simeq \left(\frac{12 }{5}\right)^{1/4} \frac{D_0^{1/2}}{\tau^{1/2} |f^{(3)}(0)|^{1/2}}\,,\qquad
\lambda_{c,1}^2 \simeq 1+ \left(\frac{12 }{5}\right)^{1/4} |f^{(3)}(0)|^{1/2} \frac{D_0^{1/2}}{\tau^{1/2} }\,.
\ee
Indeed $|f^{(3)}(0)|\,q(\lambda_{c,1})\ll1$ is again the condition (\ref{eq:weak-noise}).

White noise splits the linear instability and chaos thresholds, as observed in \cite{Schuecker_2018}, and we now work this out in the weak-noise limit. That work recognized that $|\dot C_\phi(t)|$ gives a solution of the Schr\"odinger equation (\ref{eq:Schr2model}) with $E=0$ (namely the onset of chaos) as long as $\ddot C_\phi(t=0^+)=0$. The self-consistent equation implies
\be
\tau^2 \ddot C_\phi(t=0^+) =-\partial_{C_\phi} V(C_\phi, q)\Big|_{C_\phi=q}=q-\lambda^2 \langle f^2 \rangle_q\,,
\ee
where in the last step we used (\ref{eq:wd-V-u1}). Therefore the transition to chaos for white noise occurs at
\be\label{eq:lc2white}
\lambda_{c,2}^2 \langle f^2 \rangle_q=q\,.
\ee
The Cauchy--Schwarz inequality gives $\lambda_{c,2}\ge \lambda_{c,1}$, with equality only when $f(\phi)$ is linear.

At weak noise, solving (\ref{eq:wd-white-balance2}) together with (\ref{eq:lc2white}) gives
\be\label{eq:q-lambdac2-white}
q(\lambda_{c,2})\simeq 12^{1/4}\frac{D_0^{1/2}}{\tau^{1/2} |f^{(3)}(0)|^{1/2}}\,,\qquad
\lambda_{c,2}^2 \simeq 1+12^{1/4} |f^{(3)}(0)|^{1/2} \frac{D_0^{1/2}}{\tau^{1/2} }\,.
\ee

The correlation time follows from the energy balance. Substituting the quartic expansion \eqref{eq:wd-V-u0} into \eqref{eq:wd-white-balance} gives
\be\label{eq:kappa-white-weak}
\tau^2\kappa^2 \simeq \frac{[f^{(3)}(0)]^2}{12}q^2+\frac{D_0^2}{\tau^2 q^2}\,.
\ee
At fixed $D_0$, the right-hand side is minimal, so $\tau_\textrm{corr}=1/\kappa$ is maximal, at $q=q(\lambda_{c,2})$, i.e.\ at the chaos threshold. Note that $q$ increases monotonically with $\lambda$, so this is a maximum as a function of the coupling. The value at the maximum is
\be\label{eq:taucorrexp}
\tau_{\textrm{corr,max}}= 3^{1/4}\, \frac{\tau^{3/2}}{|f^{(3)}(0)|^{1/2} D_0^{1/2}}\,.
\ee
Higher-order corrections are of relative order $D_0^{1/2}$ and can shift the maximum away from $\lambda_{c,2}$ at finite noise strength. The existence of a maximum was noted numerically in \cite{Schuecker_2018}, who also recognized its importance for optimal operation of the RNN.

We conclude that the correlation time no longer diverges at the instability or chaos thresholds in the presence of white noise. Motivated by the second-order phase transition of the Ising model in an external field, we define the noise exponent through the maximal correlation time,
\be\label{eq:nuxi}
\tau_{\textrm{corr,max}}(D_0)\equiv\max_\lambda \tau_\textrm{corr}(\lambda;D_0)\sim D_0^{-\nu_\xi}\,,\qquad D_0 \to 0\,.
\ee
For white noise, $\nu_\xi=1/2$.

We now argue that the same behavior holds for colored noise. As $D_0\to0$, the maximal correlation time $\tau_{\textrm{corr,max}}$ grows while the noise correlation time $\tau_n$ stays fixed. Once $\tau_\textrm{corr}\gg\tau_n$, the network integrates the noise over times much longer than $\tau_n$, so the noise acts as white noise of strength $D_0=\tfrac12\int dt\,G_\xi(t)$, apart from a short-time core of size $\sim\tau_n$ around $t=0$ that affects a fraction $\sim\tau_n/\tau_\textrm{corr}$ of the correlator. Hence we expect (\ref{eq:taucorrexp}) and $\nu_\xi=1/2$ to hold for general colored noise, with $D_0$ the only property of the noise that enters at leading order. Using (\ref{eq:taucorrexp}), the condition $\tau_{\textrm{corr,max}}\gg\tau_n$ reads
\be\label{eq:weakcolored}
D_0 \ll \frac{1}{|f^{(3)}(0)|}\,\frac{\tau^3}{\tau_n^2}\,.
\ee
Together with (\ref{eq:weak-noise}), the weak-noise regime for colored noise is $|f^{(3)}(0)|\,D_0/\tau\ll\min(1,\tau^2/\tau_n^2)$: for $\tau_n\lesssim\tau$ the condition (\ref{eq:weak-noise}) suffices, while for slower noise (\ref{eq:weakcolored}) is more restrictive. We verify this expectation numerically in Sec.~\ref{sec:comparison}; in particular, the collapse of the curves of $\tau_{\textrm{corr,max}}$ versus $D_0$ for different $\tau_n$ is expected to fail once $\tau_{\textrm{corr,max}}\sim\tau_n$. This universality of the cutoff exponent is one of our main results.

The condition (\ref{eq:weakcolored}) fails in the limit of quenched internal noise, $\tau_n\to\infty$ at fixed $D_\xi$, where the crossover scale in $D_0$ vanishes; this case behaves differently and is treated next.
 
\subsection{Quenched noise}\label{subsec:quenchednoise} 
 
For static random inputs $G_\xi(t)=D_q$, the correlation at late times $C_\phi(t) \to C_\infty$ determined by (\ref{eq:Cinfty}), with $q=C_\phi(0)$ fixed by the energy conservation equation (\ref{eq:Energyquenched}). 
 
For weak coupling $\lambda \ll 1$, the solution is the fixed point 
\be
C_\phi(t)\equiv q=C_\infty\,,
\ee
and (\ref{eq:Energyquenched}) fixes $q$ via the relation
\be
q-\lambda^2\langle f^2\rangle_q=D_q\,.
\ee
The next step is to expand in $q-C_\infty \ll 1$ in order to determine the coupling at which the static solution branches into a time-dependent one. At first order (\ref{eq:Cinfty}) gives
\be
q- \lambda^2 \langle f^2 \rangle_q \approx \left[ 1- \lambda^2 \langle (f')^2 \rangle_q \right] (q-C_\infty)\,.
\ee
Replacing this into the expansion (\ref{eq:wd-V-u1}) for $C_\phi=C_\infty$ around $q$ gives the energy condition at second order
\be
-V(q,q)+V(C_\infty, q)- D_q(q-C_\infty) \approx \frac{1}{2} \left[ 1- \lambda^2 \langle (f')^2 \rangle_q \right] (q-C_\infty)^2\,.
\ee
The time-dependent branch then emerges from the static one when
\begin{equation}\label{eq:lambdastar}
	\lambda_*^2\langle f'^2\rangle_{q_*}=1,\qquad q_*-\lambda_*^2\langle f^2\rangle_{q_*}=D_q .
\end{equation}
The first condition is also the instability of the fixed point, since the
spectral radius of its Jacobian is $\lambda\langle f'^2\rangle^{1/2}$. The bifurcation is continuous.

The zero mode (\ref{eq:wd-zero-mode}) is again odd and has finite norm; it corresponds to an excited state of the Schr\"odinger problem with $E=0$.  Every time-dependent
solution is therefore chaotic, and there is a single threshold
\be
\lambda_{c,1}=\lambda_{c,2}=\lambda_*\,.
\ee

We are now ready to determine the scalings in the weak disorder limit. Taking $D_q$ small in (\ref{eq:lambdastar}), and expanding for small $q_*$,
\begin{equation}
D_q=\frac{q_*\langle f'^2\rangle-\langle f^2\rangle}{\langle f'^2\rangle}
\simeq\frac{f^{(3)}(0)^2}3q_*^3\,,\,\lambda_*^2-1 \simeq (3 |f^{(3)}(0)| D_q)^{1/3}\,.
\end{equation}
In the present quenched noise limit ($\tau_n \to \infty$ with other parameters fixed), we just argued that $C_\phi(t)$ tends to a constant $C_\infty$ at large times. It is important to stress the order of limits: for fixed $\tau_n$ as $D_0 \to 0$ we will always reach a regime where $\tau_\textrm{corr, max}$ near the chaos transition becomes larger than $\tau_n$ and hence dominates the network decay time. However, this does not happen taking $\tau_n \to \infty$ first. In this case, it is convenient to use a different notion of correlation time than in the case of colored noise with $\tau_n$ fixed, namely the time decay constant of  $\delta C(t)=C_\phi(t)-C_\infty$. Subtracting the stationary equation $C_\infty=\lambda^2C_f(C_\infty,q)+D_q$ from the self-consistent equation and expanding to first order in $(\delta C)$, at fixed (q), gives
\be
-\tau^2  \delta \ddot C(t)+ \left[ 1- \lambda^2 C_{f'}(C_\infty, q)\right] \delta C(t) =0\,.
\ee
Thus
\be
\tau_\textrm{corr}= \frac{\tau}{\sqrt{1-\lambda^2 C_{f'}(C_\infty ,q)}}\,.
\ee
At weak disorder and for $\lambda$ just above $\lambda_*$, this gives
\be
\tau_\textrm{corr}^2 \simeq \frac{\tau^2}{(\lambda_*^2-1)\lambda_* (\lambda- \lambda_*)}\;.
\ee
The main conclusion of this analysis is that a static input disorder changes the fixed point value $\lambda_*=\lambda_{c,1}=\lambda_{c,2}$ but does not cut off the divergence in the correlation time. 

\subsection{Chaos and solitons}\label{subsec:solitons}

Our previous discussion of the phase transition and linear and chaos instabilities was based on perturbative expansions. Now we will improve on this, by establishing a detailed map with the theory of solitons, which will help shed light on various aspects of the dynamics of RNNs near the chaos transition.

As we established before, near the chaos transition and at weak disorder, $C_\phi(t)$ is small. Let us then perform the Mehler expansion including the first nonlinear term,
\be
C_f(t) \approx \langle f'\rangle_q^2 C_\phi(t) + \frac{1}{3!} \langle f^{(3)} \rangle_q^2 C_\phi(t)^3\,. 
\ee
Defining 
\be
\lambda_3(q) = \frac{\lambda^2}{3!} \frac{\langle f^{(3)} \rangle_q^2}{\tau^2}\,,
\ee
the self-consistent equation becomes
\be
- \ddot C_\phi(t) + \kappa(q)^2 C_\phi(t) - \lambda_3(q) C_\phi(t)^3= \frac{G_\xi(t)}{\tau^2}\,.
\ee
We recall that $\kappa(q)$ is defined by (\ref{eq:kappa}).

In the noiseless case, the energy conservation equation gives
\be\label{eq:energycons}
\frac{1}{2} \dot C_\phi(t)^2 - \frac{1}{2} \kappa(q)^2 C_\phi(t)^2 + \frac{\lambda_3(q)}{4} C_\phi(t)^4=0\,.
\ee
This corresponds to a particle that moves in the potential
\be\label{eq:Vquartic}
V=- \frac{1}{2} \kappa(q)^2 C_\phi(t)^2 + \frac{\lambda_3(q)}{4} C_\phi(t)^4\,.
\ee
The potential is self-consistent because $\kappa$ and $\lambda_3$ depend on $q=C_\phi(0)$.

The motion of the particle in this inverted potential is a well studied problem in classical and quantum mechanics and in field theory. The RNN that we are presently studying adds the new aspect regarding the coefficients $\kappa$ and $\lambda_3$ depending on the initial particle position $q$. Crucially this system admits, besides  the trivial solution $C_\phi=0$, a soliton solution
\be\label{eq:soliton1}
C_\phi^{(0)}(t) =  \sqrt{\frac{2 \kappa^2}{\lambda_3}}\, \textrm{sech}(\kappa t)\;,\;q=\sqrt{2} \frac{\kappa}{ \sqrt{\lambda_3}}\,.
\ee
The second relation between $q$ and $\kappa/\sqrt{\lambda_3}$ is the self-consistent condition for $q$; it assumes that $ q \neq 0$. When it cannot be solved, the solution is $q=0$.
Therefore, the chaotic state for the noiseless RNN is characterized by a 2-point function $C_\phi(t)$ that is the soliton associated to the quartic potential (\ref{eq:Vquartic}).

The corresponding Schr\"odinger problem (\ref{eq:Schr2model}), (\ref{eq:Schr2model2}) for the fluctuations around the solitonic solution is characterized by the P\"oschl--Teller potential
\be\label{eq:WPT}
H=-\tau^2\frac{d^2}{dt^2}+ W(t)\;,\;W(t)=\tau^2\kappa^2 \left( 1- 6\, \textrm{sech}^2 \left(\kappa t \right) \right)\,,
\ee
whose spectrum is known. It has a ground state
\be\label{eq:psib}
\psi_b(t) =\textrm{sech}^2 \left(\kappa t \right)\;,\;H \psi_b(t) =-3 \tau^2 \kappa^2 \psi_b(t)\,.
\ee
This is a bound state and corresponds to an unstable direction giving rise to the chaotic dynamics of the RNN. Furthermore, it has a normalizable zero mode due to time-translation invariance (it translates the center of $C_\phi^{(0)}(t)$)
\be
\psi_0(t)=\dot C_\phi^{(0)}(t)\;,\;H \psi_0(t)=0\,,
\ee
and then it has a continuum of states.

This connection with solitons allows to obtain various explicit results. For the noiseless case it gives an analytic expression (\ref{eq:soliton1}) for the RNN 2-point function, as well as a result for the leading unstable fluctuation (\ref{eq:psib}). In particular, this establishes that near the chaos transition the Lyapunov exponent is determined by the correlation time
\be
\Lambda_L \approx \frac{3}{2} \kappa^2 \tau=\frac{3}{2} \frac{\tau}{\tau_\textrm{corr}^2}\,.
\ee

The formulation is also useful in the presence of input noise. The simplest situation arises with white noise; then we also have the energy conservation equation (\ref{eq:energycons}) for $t>0$, but there is a discontinuity
\be\label{eq:discdCphi}
\dot C_\phi(0^+)-\dot C_\phi(0^-)=-\frac{2 D_0}{\tau^2}\,,
\ee
and so $C_\phi(t)$ has a cusp at $t=0$. We look for a soliton solution as in the noiseless case, but with a time shift and a cusp:
\be\label{eq:whitesoliton}
C_\phi(t) = \sqrt{\frac{2 \kappa^2}{\lambda_3}} \,\textrm{sech}\left(\kappa (|t|+t_0)\right)\,.
\ee
Replacing this into (\ref{eq:discdCphi}) determines the time shift
\be
\textrm{sech}(\kappa t_0) \tanh(\kappa t_0)=\frac{D_0}{\tau^2\kappa^2} \sqrt{\frac{\lambda_3}{2}}\,.
\ee
Note that this has solution only for sufficiently weak disorder; in particular, as $D_0 \to 0$,
\be
\kappa t_0\simeq \frac{D_0}{\tau^2\kappa^2} \sqrt{\frac{\lambda_3}{2}}\,.
\ee
The effect of the internal noise is then to modify the core of the soliton at time scales of order $t_0$.

The associated Schr\"odinger potential is now the even cusped function 
\be
W(t)=\tau^2\kappa^2\big(1-6\,\textrm{sech}^2[\kappa(|t|+t_0)]\big)\,.
\ee 
The spectrum can also be found, but we won't need explicit expressions.  The weak noise limit can be treated using perturbation theory in $t_0$. At first order, the ground state energy is found to change as $\delta E_0= 3\kappa^3 \tau^2 t_0$ and then
\be
\Lambda_L \tau \approx \frac{3}{2} \left(\kappa^2 \tau^2- \sqrt{\frac{\lambda_3}{2}} D_0 \right)\,.
\ee

The formalism can also be applied to a general colored noise, and various directions could be pursued, including perturbation theory around the noiseless soliton, and the inclusion of collective coordinates. But we won't develop this further here and will next present the results from simulations and theoretical predictions away from weak disorder.

\section{Results: theory and simulations}\label{sec:comparison}

All simulation results presented in this section correspond to networks of $N=4000$ neurons. Finite-size effects are analyzed separately in Appendix~\ref{app:finiteN}. We compare these simulations with the large-$N$ description. We first examine the temporal correlation of the preactivations, distinguishing a test of the correlation equation using measured output correlations from an approximate prediction obtained without measured input. We then determine the thresholds for local linear instability and chaos, and study their dependence on the activation exponent and the noise correlation time.

\subsection{Correlators}\label{subsec:corrs}

\begin{figure}[h!]
	\centering
	\includegraphics[width=1.0\textwidth]{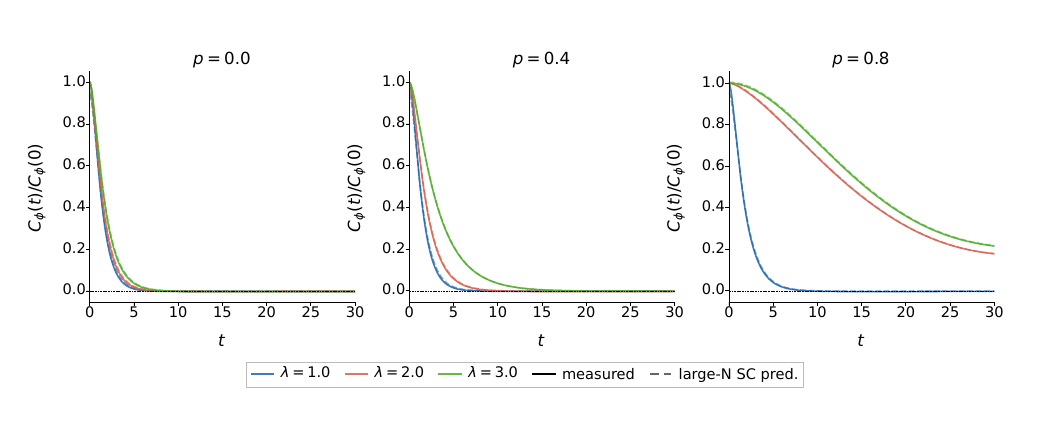}
	\caption{Normalized preactivation autocorrelation $C_\phi(t)/C_\phi(0)$ for $N=4000$, $\tau=1$, $\beta=0.6$, $\tau_n=0.5$, $D_\xi=40$, and $\lambda=1$, $2$ and $3$. The panels show $p=0$, $0.4$ and $0.8$. Solid curves are measured in the simulations; dashed curves are reconstructed from the large-$N$ correlation equation using the measured $C_f(t)$.}
	\label{fig:correlators}
\end{figure}

Figure~\ref{fig:correlators} compares the normalized autocorrelation $C_\phi(t)/C_\phi(0)$ measured in the network with the result obtained from Eq.~\eqref{eq:selfrepeat}, using the measured $C_f(t)$ as its recurrent source. The noise parameters are $\tau_n=0.5$ and $D_\xi=40$, and the three panels correspond to $p=0$, $0.4$ and $0.8$, each at couplings $\lambda=1$, $2$ and $3$. The measured and reconstructed curves are nearly indistinguishable over the displayed time interval. This comparison tests the relation between input and output correlations predicted by the large-$N$ theory, while retaining the output correlations of the finite network as input.

The temporal structure changes markedly with the activation exponent. For the saturating response $p=0$, the normalized correlation decays rapidly and depends only weakly on coupling over the range shown. At $p=0.4$, increasing $\lambda$ produces a more visible slowing of the decay. This effect is much stronger at $p=0.8$: whereas the curve at $\lambda=1$ decays rapidly, those at $\lambda=2$ and $3$ retain appreciable correlations throughout the observation window. Thus, recurrent interactions can generate a time scale much longer than the input correlation time, particularly for responses closer to the linear limit. The slowly decaying tails in the last panel should not be interpreted as evidence for a nonzero infinite-time plateau. Moreover, a long correlation time alone does not determine whether the dynamics is chaotic; that distinction requires the Lyapunov analysis below.

\begin{figure}[h!]
	\centering
	\includegraphics[width=1.0\textwidth]{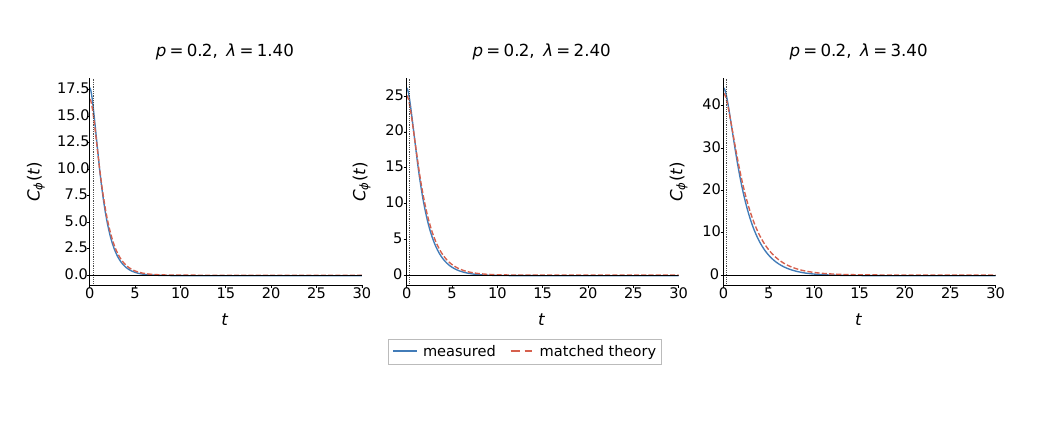}
	\caption{Unnormalized preactivation autocorrelation $C_\phi(t)$ measured in the simulations (solid curves) and predicted by matching the short-time solution to the long-time Mehler expansion through $k=3$ (dashed curves), without measured input. The parameters are $N=4000$, $p=0.2$,$\tau=1$, $\beta=0.6$, $\tau_n=0.5$, $D_\xi=40$, and $\lambda=1.4$, $2.4$ and $3.4$ from left to right. Each panel reports the predicted and measured variance $q$, the predicted intrinsic correlation time $1/\kappa$, and the matching time $t^*$.}
	\label{fig:correlators-theory}
\end{figure}

A separate test is provided by Figure~\ref{fig:correlators-theory}, where $C_\phi(t)$ is predicted by the matching procedure of Sec.~\ref{subsec:approxsolution}, without using measured correlations as input. The short-time solution is matched to the long-time Mehler expansion, retaining the terms with $k=1$ and $3$. The figure displays the unnormalized correlator for $p=0.2$ and couplings $\lambda=1.4$, $2.4$ and $3.4$, at the same noise parameters as Fig.~\ref{fig:correlators}. The predicted curves reproduce both the amplitude and the overall decay of the measured correlations, with small deviations.

The predicted solution from the matching procedure can be systematically improved by adding higher order contributions; however, our goal with this comparison has been to establish that the main dynamical properties of the 2-point correlation function are already captured by the first terms in the theoretical expansion. This agreement supports the use of matched short- and long-time expansions to describe the correlator beyond the weak-disorder regime. Together, the two figures test both the correlation equation itself and the predictive accuracy of its approximate solution.

\subsection{Phase diagram and correlation time}

\begin{figure}[h!]
	\centering
	\includegraphics[width=0.8\textwidth]{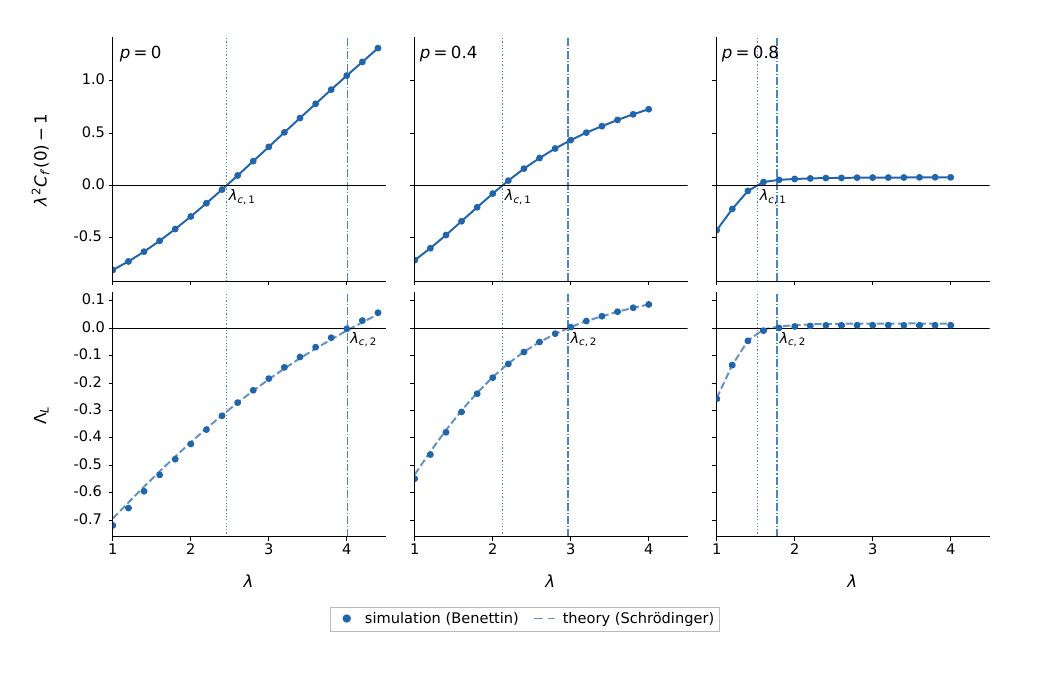}
	\caption{Determination of the local-instability and chaos thresholds for $N=4000$,$\tau=1$, $\beta=0.6$, $\tau_n=0.5$, $D_\xi=40$, and $p=0$, $0.4$ and $0.8$ from left to right. The upper panels show $\lambda^2 C_{f'}(0)-1=\lambda^2\langle(f')^2\rangle_q-1$, whose zero defines $\lambda_{c,1}$. The lower panels compare the largest Lyapunov exponent from simulations (symbols) with the prediction from the effective Schr\"odinger problem (dashed curves); its zero defines $\lambda_{c,2}$. Vertical dotted and dash-dotted lines mark $\lambda_{c,1}$ and $\lambda_{c,2}$, respectively.}
	\label{fig:lambda_determination}
\end{figure}

Figure~\ref{fig:lambda_determination} shows how the two dynamical
thresholds are extracted at $\tau_n=0.5$ and $D_\xi=40$, for
$p=0$, $0.4$ and $0.8$. The upper panels plot
$\lambda^2 C_{f'}(0)-1$, evaluated using $C_{f'}(0)$ measured
in the simulations; its zero defines
$\lambda_{c,1}$ according to Eq.~\eqref{eq:lambdac1}. The lower
panels show the largest Lyapunov exponent measured by the
Benettin procedure, together with the theoretical result from
the ground-state energy of Eq.~\eqref{eq:Schr2model},
$\Lambda_L=(\sqrt{1-E_0}-1)/\tau$, obtained using the simulated
$C_{f'}(t)$ in the Schr\"odinger potential. The close agreement
tests the large-$N$ relation between perturbation growth and the
stationary derivative correlator.

The two zero crossings are distinct. For $p=0$, local instability occurs at approximately $\lambda_{c,1}=2.5$, while the Lyapunov exponent becomes positive only near $\lambda_{c,2}=4$. Increasing the exponent to $p=0.4$ shifts the thresholds to approximately $2.1$ and $3$, and at $p=0.8$ both occur below $\lambda=2$, with a much smaller separation. Thus, three regimes can be distinguished: local linear stability below $\lambda_{c,1}$; local instability but a negative largest Lyapunov exponent between $\lambda_{c,1}$ and $\lambda_{c,2}$; and chaos above $\lambda_{c,2}$. In the intermediate regime, instantaneous local instability does not imply asymptotic exponential separation of nearby trajectories driven by the same noise history. This is consistent with the theoretical distinction between a negative minimum of the effective Schr\"odinger potential and a negative ground-state energy: the former is necessary for chaos but is not sufficient.

\begin{figure}[h!]
	\centering
	\includegraphics[width=0.6\textwidth]{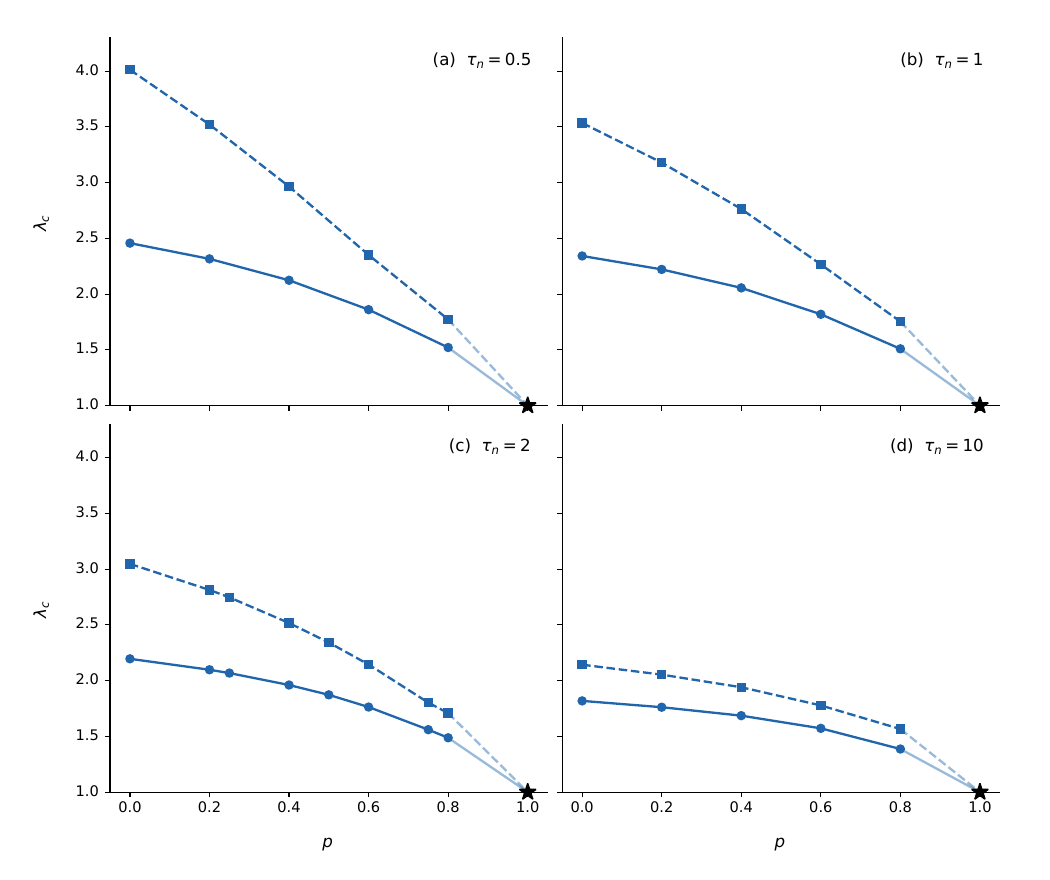}
	\caption{Local-instability threshold $\lambda_{c,1}$ (solid curves, circles) and chaos threshold $\lambda_{c,2}$ (dashed curves, squares) as functions of the activation exponent $p$, for $N=4000$,$\tau=1$, $\beta=0.6$. Panels (a)--(d) correspond to $\tau_n=0.5$, $1$, $2$ and $10$, respectively, with fixed integrated noise strength $D_0=D_\xi\tau_n=20$. Both increasing $p$ and increasing $\tau_n$ reduce the separation between the thresholds. Stars at $p=1$, $\lambda=1$ mark the large-$N$ stability boundary of the linear model; the connections to these stars indicate the linear endpoint.}
	\label{fig:phase_diagram}
\end{figure}

Figure~\ref{fig:phase_diagram} extends this comparison to noise correlation times $\tau_n=0.5$, $1$, $2$ and $10$. Here the integrated noise strength is held fixed, $D_0=D_\xi\tau_n=20$. Two properties are clear. First, both thresholds decrease as $p$ increases, and the interval $\lambda_{c,1}<\lambda<\lambda_{c,2}$ narrows. For example, at $\tau_n=0.5$ the gap decreases from approximately $1.5$ at $p=0$ to about $0.25$ at $p=0.8$. Second, increasing the noise correlation time lowers both thresholds and brings them closer together. At $p=0$, increasing $\tau_n$ from $0.5$ to $10$ moves the local-instability threshold from approximately $2.5$ to $1.8$ and the chaos threshold from approximately $4$ to $2.1$. Slowly varying input therefore leaves a substantially smaller intermediate regime along this fixed-$D_0$ scan. This is qualitatively consistent with the coincidence of the two thresholds for strictly quenched input established in Sec.~\ref{subsec:quenchednoise}.

The star at $p=1$, $\lambda=1$ marks the stability boundary of the linear model. At this endpoint $f(\phi)=\phi$, and the local and asymptotic stability criteria coincide in the large-$N$ limit. This point is a limiting reference for the nonlinear phase diagrams: beyond the linear stability boundary, the linear model has no nonlinear mechanism to bound the activity and does not represent a stationary chaotic phase. For the nonlinear responses $p<1$, the separation of the thresholds demonstrates that the temporal structure of the noise and the form of the activation jointly control whether local instability develops into chaos.

The correlation time, defined in Eq. (\ref{eq:taucorr}), is an important observable in the dynamics of RNNs, as we studied in detail in Sec. \ref{sec:weak-disorder}. We will now analyze its behavior numerically, using the solution obtained by matching the short and long time behaviors, developed in Sec. \ref{subsec:approxsolution}. As shown in Figure~\ref{fig:correlators-theory}, already the solution that keeps the leading short-time effects (\ref{eq:Chatshort}) and the first two terms ($k=1, k=3$) in the Mehler expansion (\ref{eq:mehler}), gives a very good agreement with simulations.  This approximate solution is well suited for studying the correlation time near the chaos transition, where we expect it to become a maximum and much larger than the basic time constant $\tau$.

\begin{figure}[h!]
	\centering
	\includegraphics[width=0.8\textwidth]{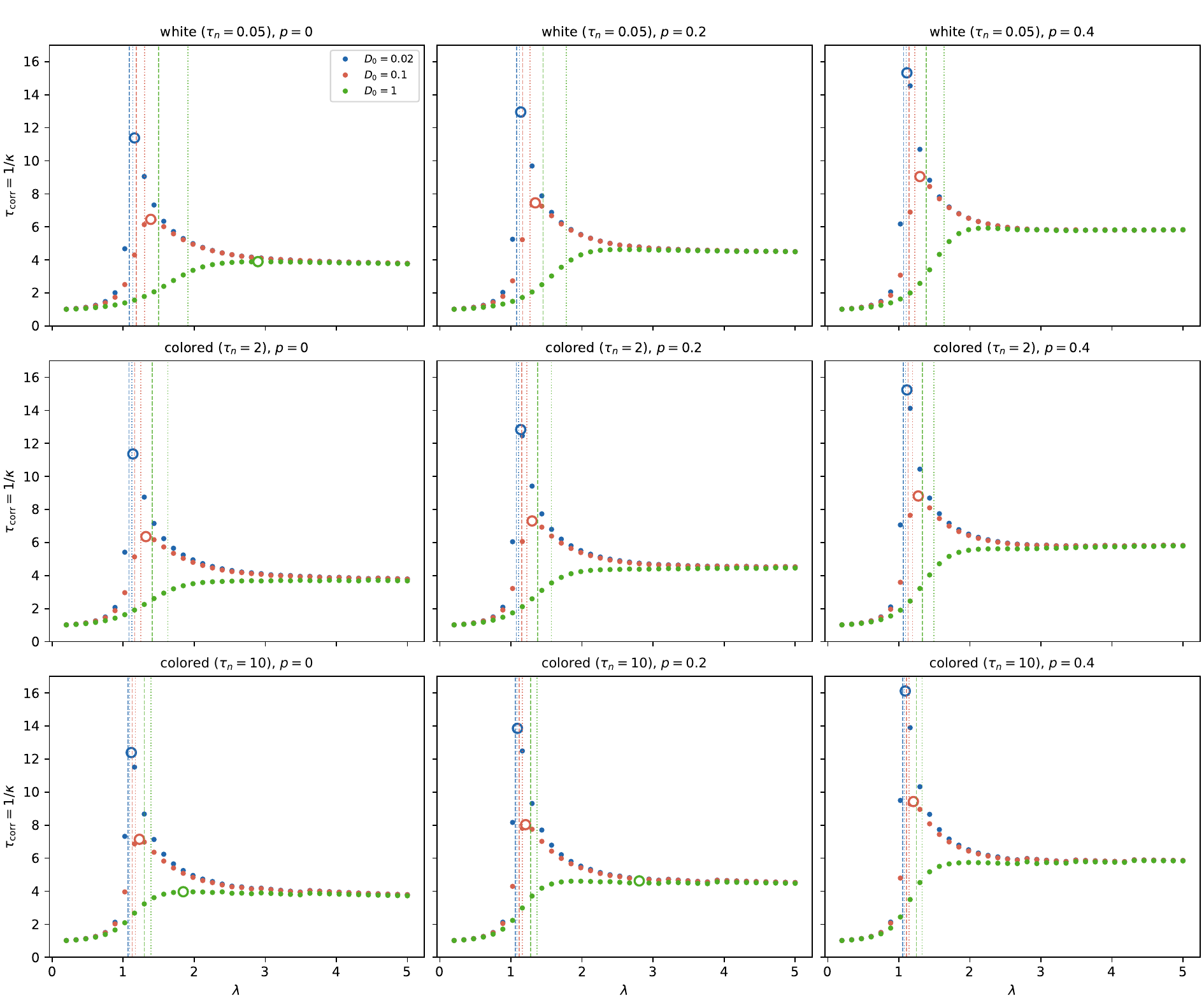}
	\caption{Correlation time vs coupling $\lambda$ near the limit of white noise (first row panels) and for colored noise (second and third row panels). Three different values of $p$ are also considered. The maximum occurs at sufficiently small internal noise strength.}
	\label{fig:taucorrlambda}
\end{figure}

First, in Figure~\ref{fig:taucorrlambda} we present results for the correlation time for varying coupling. This is done for small $\tau_n$ (close to the white limit) and for $\tau_n \sim \tau$. We also explore different values of $p$ and of $D_0=D_\xi \tau_n$. For white and colored noise, and for the values of $p$ explored, we find a maximum of $\tau_\textrm{corr}$ that occurs once $D_0$ is sufficiently small. At the smallest values that we explored, the maximum occurs at $\lambda_{c,2}$, consistent with our theoretical analysis of the weak disorder limit. As the disorder strength is increased, the maximum first moves to $\lambda> \lambda_{c,2}$ and then disappears.

\begin{figure}[h!]
	\centering
	\includegraphics[width=0.8\textwidth]{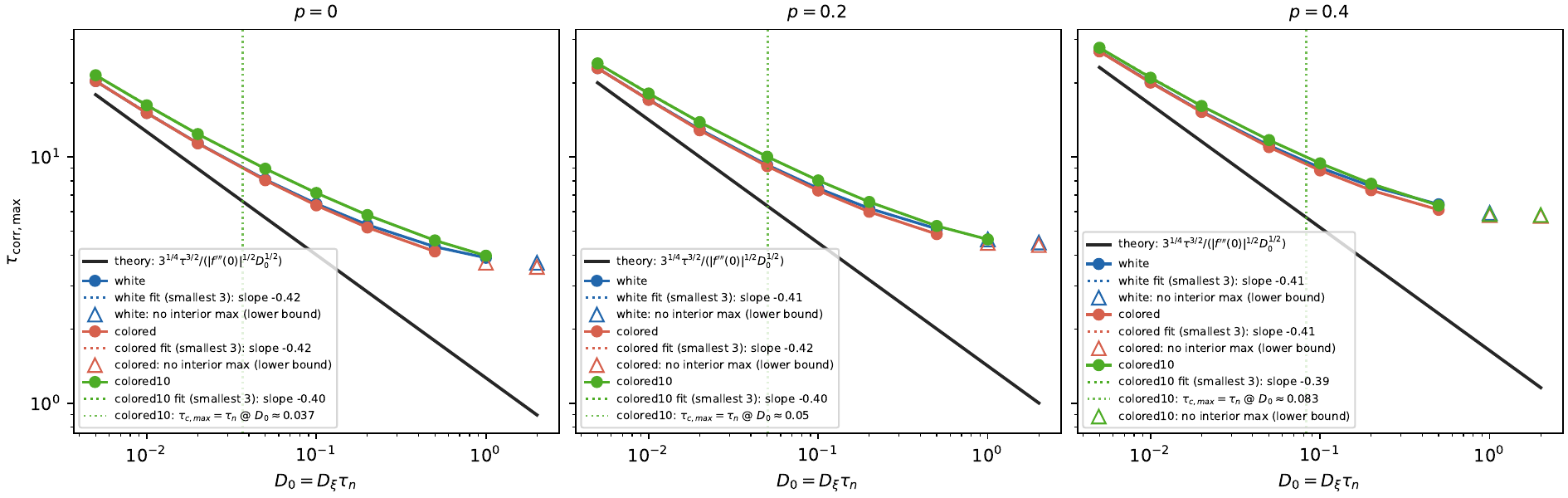}
	\caption{$\tau_\textrm{corr, max}$ as a function of $D_0$, together with the theory curve \ref{eq:taucorrexp}). The agreement improves as $D_0 \to 0$, as expected. We show the cross-over scale $\tau_\textrm{corr, max} \sim \tau_n$.}
	\label{fig:taucorrmaxD0}
\end{figure}

Next, in Figure~\ref{fig:taucorrmaxD0}, we plot the maximum value for $\tau_\textrm{corr}$ as a function of $D_0$ as well as the theoretical result (\ref{eq:taucorrexp}). Both white and colored noise curves approach the theoretical one for sufficiently small $D_0$ and verify the critical exponent $\nu_\xi=1/2$. This provides a numerical validation for our argument that near the chaos transition and $D_0 \to 0$, colored noise should behave as white noise. Deviations arise once $\tau_\textrm{corr, max}$ becomes comparable to $\tau_n$.

\section{Four-point function and dimension of participation}\label{sec:dpr}

The preceding sections characterized temporal correlations, local instability and the onset of chaos. We now ask how recurrent coupling organizes activity across neurons. At each frequency, the participation dimension quantifies the effective
number of population modes contributing to the spectral power of the
output activity $f(\phi)$. Previous work has related participation dimension to the statistics of pairwise
covariances in recurrent networks \cite{dahmen2020strong,dahmen2019second,clark2023dimension}. In our previous work, we calculated correlation statistics and the
participation dimension with quenched connectivity $W$ and quenched
internal noise, in the limit $\tau_n/\tau\to\infty$
\cite{Mato:2025yof}. Nonlinear recurrent
networks driven by white internal noise, corresponding to the
opposite limit $\tau_n/\tau\to0$ were studied in \cite{shen2025covariance}.
Here we extend these results to colored noise with a finite
correlation time.

\subsection{Setup}\label{subsec:dprsetup}

 At fixed $W$, let $C_{ij}^f(\omega)$ be the
Fourier transform of $\langle f(\phi_i(t))f(\phi_j(0))\rangle_\xi$; the disorder-averaged
diagonal is $C_f(\omega)$ to leading order, as in
\eqref{eq:defCphiCfmodel}. Stationarity at fixed $W$ implies
$C_{ji}^f(\omega)=C_{ij}^f(-\omega)$, and hence
\be\label{eq:DPRtraceDPR}
D_{PR}(\omega)
=\frac{[\operatorname{Tr}C^f(\omega)]^2}
{\operatorname{Tr}[C^f(\omega)^2]}=\frac{[\sum_i C_{ii}^f(\omega)]^2}
{\sum_i C_{ii}^f(\omega)^2+
 \sum_{i\ne j}C_{ij}^f(\omega)C_{ij}^f(-\omega)}\,.
\ee
The off-diagonal term in the denominator involves a product of two noise-averaged correlators. This is why a four-point calculation is required even at leading order in $D_{PR}/N$: the contribution of each distinct pair is $O(1/N)$, but there are $O(N^2)$ pairs. We write the disorder-averaged four-point quantity associated with
this off-diagonal product as
\be\label{eq:Gammadefmodel}
\Gamma^f(\omega)=\langle \langle C_{ij}^f(\omega)\rangle_\xi 
\langle \langle C_{ij}^f(-\omega)\rangle_\xi \rangle_W \Big|_{i \neq j}\,.
\ee
Each factor $C_{ij}^f$ is a noise-averaged correlation
at fixed $W$; the outer brackets average their product over the
quenched connectivity. Then $\Gamma^f$ measures the disorder-averaged
second moment of the correlations between distinct neurons generated
by the recurrent coupling and together with $C_f(\omega)$, it determines
the leading large-$N$ expression for the frequency-resolved
participation dimension. Eq.~\eqref{eq:DPRtraceDPR} gives, wherever $C_f(\omega)\ne0$,
\be\label{eq:DPRGammaDPR}
\frac{D_{PR}(\omega)}{N}
=\frac{1}{1+N\Gamma^f(\omega)/C_f(\omega)^2}.
\ee

Calculating \eqref{eq:Gammadefmodel} requires two replicas that share $W$ but have independent noise histories, so that their correlation functions multiply before the $W$ average. (In contrast, the two copies used for the Lyapunov exponent in Sec.~\ref{subsec:chaos} instead share the same noise history). Bilocal sources $J_i^{ab}(t,t')$ with $a\ne b$ generate the required four-point function. Their quadratic generating functional contains a term proportional to $\sum_iJ_iJ_i$, which contributes at the same site, and one proportional to $(\sum_iJ_i)^2$, whose derivative at $i\ne j$ gives $\Gamma^f$. 

The off-diagonal quantity $\Gamma^f(\omega)$ in Eq.~\eqref{eq:Gammadefmodel} contains a product of two noise-averaged cross-spectra at fixed $W$. We therefore simulate $M$ independent noise histories with the same coupling matrix $W$ and, after discarding the transient, retain $L$ samples $y_{ik}^{(r)}=f(\phi_i^{(r)}(k\delta t))$ from each history $r$. Since we estimate a participation dimension at each frequency from finite segments, the abrupt endpoints of an untapered Fourier transform can cause spectral leakage between frequency bins. Before taking the transform, we multiply each segment by the same Hann window $w_k$, which tapers its endpoints to zero and reduces this leakage at the cost of broader spectral resolution. For the discrete frequencies $\omega_\ell=2\pi\ell/(L\delta t)$, we compute
\begin{align}\label{eq:DPRwindowDFT}
G_i^{(r)}(\omega_\ell)&=\sum_{k=0}^{L-1}w_k\,y_{ik}^{(r)}e^{-2\pi i\ell k/L}\,,\nonumber\\
w_k&=\frac{1}{2}\left[1-\cos\!\left(\frac{2\pi k}{L-1}\right)\right],\qquad
s=\frac{\delta t}{\sum_{k=0}^{L-1}w_k^2}=\frac{8\delta t}{3(L-1)}\,.
\end{align}
The window in Eq.~\eqref{eq:DPRwindowDFT} also changes the power of the Fourier coefficients: $\sum_k w_k^2=3(L-1)/8$, compared with $L$ for a rectangular window. The factor $s$ corrects for this change in the periodogram $s|G_i^{(r)}|^2$ and, correspondingly, enters $\widehat C_f$ once and $\widehat\Gamma^f$ twice in Eq.~\eqref{eq:DPRestimators}. It replaces $\delta t/L$ by a factor $8L/[3(L-1)]$ larger; it does not undo the finite-frequency smoothing caused by the window. No sample mean or linear trend is subtracted in these simulations. In the symmetric stationary state the ensemble mean of $f(\phi_i)$ vanishes, while the finite-window fluctuations, including those at $\omega_0=0$, contribute to the measured spectrum.

At each frequency we define the site sums
\be\label{eq:DPRnumericalSums}
P_r(\omega)=\sum_i|G_i^{(r)}(\omega)|^2\,,\,
A_{rs}(\omega)=\sum_iG_i^{(r)}(\omega)G_i^{(s)}(\omega)^*\,,\,
B_{rs}(\omega)=\sum_i|G_i^{(r)}(\omega)|^2|G_i^{(s)}(\omega)|^2\,.
\ee
Here $sG_i^{(r)}G_j^{(r)*}$ estimates $C_{ij}^f(\omega)$ from history $r$. Expanding $|A_{rs}|^2$ sums the products of this cross-spectrum from histories $r$ and $s$ over all neuron pairs $(i,j)$; $B_{rs}$ removes precisely the $i=j$ contributions. Hence, at fixed $W$,
\be\label{eq:DPRestimators}
\widehat C_f(\omega)=\frac{s}{NM}\sum_r P_r(\omega)\,,\qquad
\widehat\Gamma^f(\omega)=\frac{s^2}{N(N-1)M(M-1)}
\sum_{r\ne s}\big[|A_{rs}(\omega)|^2-B_{rs}(\omega)\big] \,.
\ee
The two exclusions have different roles: $i\ne j$ selects the off-diagonal covariance required by $\Gamma^f$, whereas $r\ne s$ removes the finite-$M$ self-pair bias in a product of noise averages as required by \eqref{eq:Gammadefmodel}. 

For a direct check on the finite-sample estimator of Eq.~\eqref{eq:DPRtraceDPR}, we also form the U-statistics
\be\label{eq:DPRdirectU}
\widehat T_U=\frac{s^2}{M(M-1)}\sum_{r\ne s}P_rP_s\,,\,
\widehat Q_U=\frac{s^2}{M(M-1)}\sum_{r\ne s}|A_{rs}|^2\,,\,
\frac{D_{PR,U}(\omega)}{N}=\frac{\widehat T_U}{N\widehat Q_U}\,.
\ee
Both ratios are evaluated separately for each $W$, and their means and standard deviations are then taken over the independent $W$ realizations. The factors $s$ cancel in either ratio, but they are needed to report $\widehat C_f$ and $\widehat\Gamma^f$ with their correct spectral normalization. The independent-noise histories here differ from the same-noise copies used to estimate the Lyapunov exponent below.

\subsection{Large $N$ results}\label{subsec:dprresult}

We present the theoretical large $N$ calculation of the 4-point function and dimension of participation in Appendix \ref{app:fourpoint} using collective fields and replicas.
Working at the same self-consistent saddle discussed in Sec.~\ref{sec:largeN}, and introducing the effective coupling
\be
\lambda_{eff}= \lambda \langle f'(\phi(t=0)) \rangle_q\,,
\ee
the on-shell effective action at quadratic order in the sources gives, in frequency space,
\be\label{eq:Gammafinalmodel}
\Gamma^f(\omega)=\frac{1}{N}C_f(\omega)^2\left[\frac{(1+\tau^2\omega^2)^2}{(1+\tau^2\omega^2-\lambda_{eff}^2)^2}-1\right]+O(N^{-2})\,,
\ee
The frequency-dependent participation ratio then takes the  simple closed form
\be\label{eq:DPRomega}
\frac{D_{PR}(\omega)}{N}= \left(1- \frac{\lambda_{eff}^2}{1+\tau^2 \omega^2} \right)^2\,.
\ee
This agrees with the white-noise result of \cite{shen2025covariance}. The effect from colored noise enters via the self-consistent values of  $C_f,\,C_\phi$ and hence $\lambda_{eff}$ from Sec.~\ref{sec:largeN}. We discuss the strong-coupling scaling of $\lambda_{eff}$, and hence of $D_{PR}(\omega)$, together with the numerical comparison, in Sec.~\ref{sec:comparison}.

The spectral participation dimension can be related directly to the intrinsic correlation time. Eqs.~\eqref{eq:kappa} and \eqref{eq:taucorr} imply 
\be
\lambda_{eff}^2=1-\tau^2/\tau_\textrm{corr}^2\,.
\ee
 Substituting this identity into \eqref{eq:DPRomega} gives
\be\label{eq:DPR-kappa}
\frac{D_{PR}(\omega)}{N}=\left(\frac{\tau^2\omega^2+\tau^2/\tau_\textrm{corr}^2}{1+\tau^2\omega^2}\right)^2\,,\qquad
\frac{D_{PR}(0)}{N}=\tau^4\kappa^4=\left(\frac{\tau}{\tau_\textrm{corr}}\right)^4\,.
\ee
This relation is an important result of the large-$N$ theory. It connects the temporal analysis of Secs.~\ref{sec:largeN}--\ref{sec:weak-disorder} with the four-point calculation of the present section. A longer intrinsic correlation time corresponds to a smaller zero-frequency participation fraction, and the transition from low to high frequency dominance in the dimension of participation occurs at $\omega \sim 1/\tau_\textrm{corr}$.

In Sec. \ref{sec:weak-disorder} we showed that for weak disorder strength $D_0 \to 0$ (both for white and colored noise), the correlation time attains a maximum near the transition to chaos.
This implies that at weak disorder the transition to chaos is detected as a minimum for the dimension of participation at zero frequency,
\be
\left.\frac{D_{PR}(0)}{N}\right|_{\lambda_{c,2}}= \frac{(f^{(3)}(0))^2 D_0^2}{3\tau^2}\,.
\ee
This feature disappears at larger internal disorder.

\subsection{Comparison with simulations}\label{subsec:dprsimulation}

We compare the output spectral participation dimension with the large-$N$ prediction for $p=0$ and $p=0.5$, using $N=4000$, $\tau=1$, $\beta=0.6$, $D_\xi=10$, and $\tau_n=2$. For the theoretical results we use (\ref{eq:DPRomega}), computed in terms of the correlators measured in simulations. The final spectral estimates follow Sec.~\ref{subsec:dprsetup}: $M=100$ independent noise histories at fixed $W$, with Hann-windowed segments of $L=4096$ samples and $\delta t=0.1$. Thus the segment duration is $T_{\rm seg}=409.6$, the frequency spacing is $\Delta\omega=2\pi/T_{\rm seg}$, and the Nyquist frequency is $\omega_{\rm Nyq}=\pi/\delta t$. Ratios are formed separately for the coupling seeds $W=0,\ldots,4$ before averaging. Error bars and shaded bands report the standard deviation across these matrices, rather than the uncertainty of the disorder mean.

\begin{figure}[h!]
\centering
\includegraphics[width=0.75\textwidth]{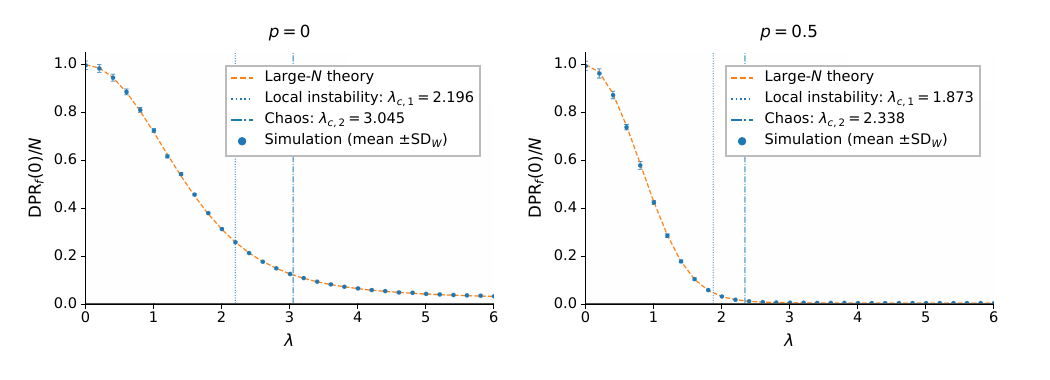}
\caption{Zero-frequency output spectral participation fraction versus coupling for $p=0$ (left panel) and $p=0.5$ (right panel), at $N=4000$. Symbols are simulation means, with one standard deviation across five $W$ realizations; dashed orange curves are the large-$N$ theoretical comparison. Dotted and dash-dotted vertical lines mark the local-instability and chaos thresholds, respectively. Parameters: $\tau=1$, $\beta=0.6$, $D_\xi=10$, $\tau_n=2$, $M=100$, $L=4096$, $\delta t=0.1$, and Hann window.}
\label{fig:dprlambda}
\end{figure}

Figure~\ref{fig:dprlambda} compares the measured zero-frequency output dimension with the theoretical curves. Both activation functions show a progressive reduction of $D_{PR}(0)/N$ as the recurrent coupling increases. The reduction is appreciably stronger for $p=0.5$; the output is already concentrated in a small participation fraction in the intermediate, locally unstable nonchaotic regime. The independently determined thresholds are $(\lambda_{c,1},\lambda_{c,2})=(2.196,3.045)$ for $p=0$, and $(1.873,2.338)$ for $p=0.5$. Low-dimensional spectral activity occurs before the largest Lyapunov exponent becomes positive, and the dimension continues to decrease within the chaotic regime.

\begin{figure}[h!]
\centering
\includegraphics[width=\textwidth]{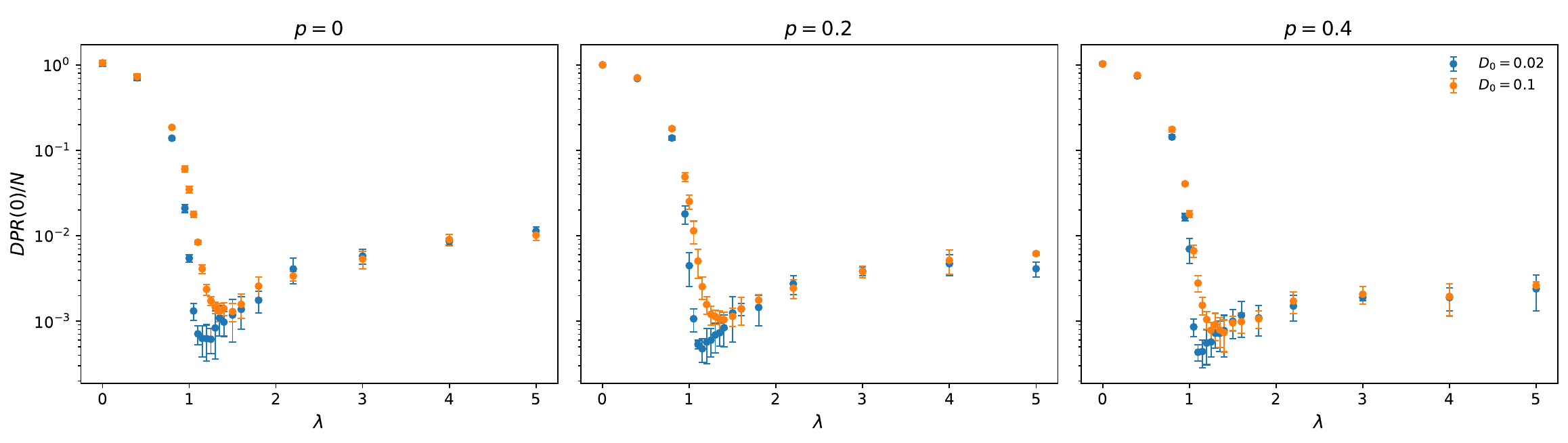}

\medskip
\includegraphics[width=\textwidth]{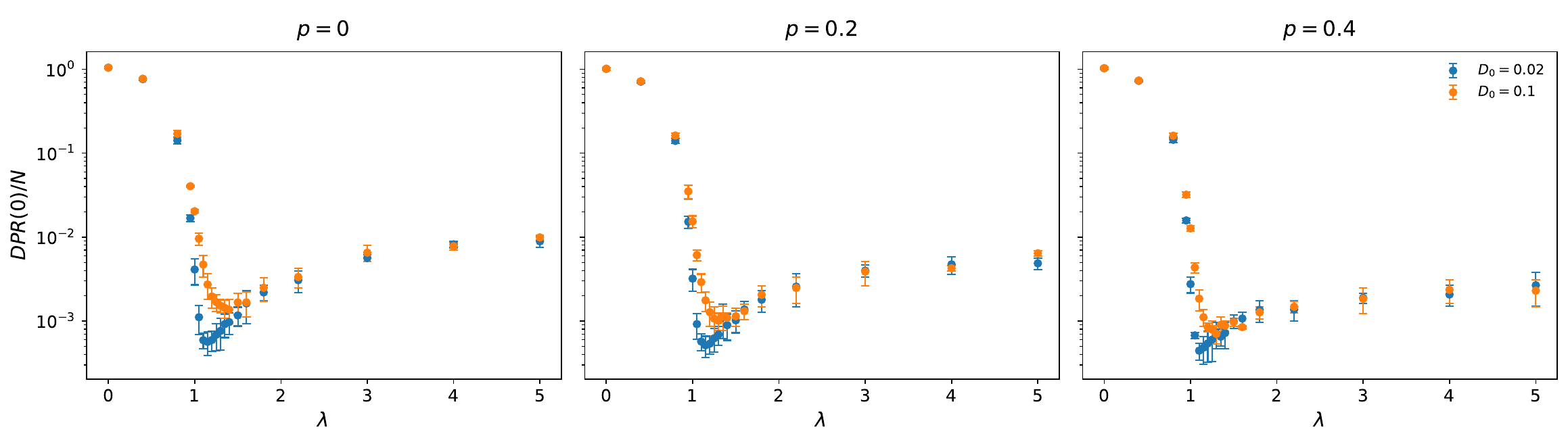}
\caption{Zero-frequency participation fraction $D_{PR}(0)/N$ versus recurrent coupling from simulations with $N=4000$. Top row: nearly white noise, $\tau_n=0.05$. Bottom row: colored noise, $\tau_n=2$. In both rows, the activation exponents are $p=0$, $0.2$, and $0.4$ from left to right. Blue and orange symbols correspond to integrated noise intensities $D_0=D_\xi\tau_n=0.02$ and $0.1$, respectively. The vertical axes are logarithmic. All six cases exhibit a minimum followed by an increase with coupling, with a deeper minimum at the smaller noise intensity.
 Simulations use Euler–Maruyama with an xi-first OU update; \(\delta t=0.005\) and \(L=81920\) for \(\tau_n=0.05\), while \(\delta t=0.1\) and \(L=4096\) for \(\tau_n=2\). The spectral estimator uses a Hann window.}
\label{fig:dprminimum}
\end{figure}

The additional simulations in Fig.~\ref{fig:dprminimum} probe the weak-noise regime for both nearly white and colored input, with $N=4000$ and integrated noise intensities $D_0=D_\xi\tau_n=0.02$ and $0.1$. For nearly white noise, $\tau_n=0.05$, the three activation exponents $p=0$, $0.2$, and $0.4$ all display a sharp decrease of the zero-frequency participation fraction near $\lambda=1$, followed by a minimum and a subsequent increase. The minimum is deeper for the smaller noise intensity. For colored noise, $\tau_n=2$, the same qualitative nonmonotonic behavior is observed for all three activation exponents. These results are consistent with the predicted relation $D_{PR}(0)/N=(\tau/\tau_\textrm{corr})^4$: a maximum of the intrinsic correlation time produces a minimum of the zero-frequency participation fraction. The simulations support the persistence of this minimum at finite noise correlation time.

\begin{figure}[h!]
\centering
\includegraphics[width=0.75\textwidth]{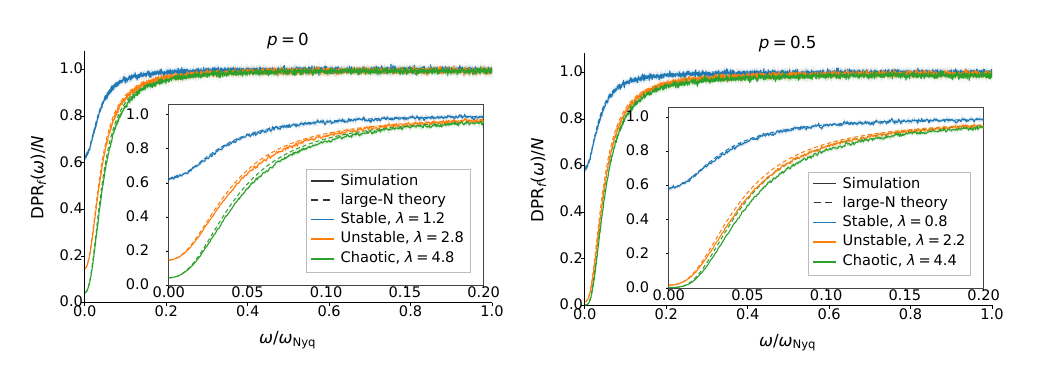}
\caption{Output spectral participation fraction across representative dynamical regimes for $p=0$ (left panel: $\lambda=1.2,2.8,4.8$) and $p=0.5$ (right panel: $\lambda=0.8,2.2,4.4$). Solid curves are simulated means and shaded bands are one standard deviation across five $W$ realizations; dashed curves are the large-$N$ theoretical comparison. The horizontal axis is $\omega/\omega_{\rm Nyq}$, with $\omega_{\rm Nyq}=\pi/\delta t$. Parameters and Hann-window estimation are those of Fig.~\ref{fig:dprlambda}. ``Unstable'' denotes the locally unstable nonchaotic regime.}
\label{fig:dprspectra}
\end{figure}

The frequency-resolved comparison in Figure~\ref{fig:dprspectra} samples one coupling in each regime. At $p=0$ the selected couplings are $1.2$, $2.8$, and $4.8$; at $p=0.5$ they are $0.8$, $2.2$, and $4.4$. In both cases the curves differ most strongly near zero frequency and, as expected, approach a participation fraction close to one at higher frequencies. The theoretical curves reproduce the main crossover and its dependence on coupling in all three regimes.

The fixed-frequency coupling sweeps provide a complementary check. At $\omega/\omega_{\rm Nyq}=0.25$ and $0.50$, the supplied simulations remain close to full participation throughout the coupling range, with a much weaker variation than at zero frequency. For the present $\delta t$ and $\tau$, these frequencies correspond to $\omega\tau\simeq7.85$ and $15.71$, where the recurrent correction in Eq.~\eqref{eq:DPRomega} is small. The recovery of a large participation fraction at these frequencies occurs in the locally unstable and chaotic regimes as well as in the stable regime.

The coupling and frequency sweeps show that the output covariance can be described by the same effective-gain framework across the three dynamical regimes. The observations support a continuous reorganization of low-frequency output correlations with recurrent coupling.

\section{Discussion}\label{sec:discussion}

We studied large random recurrent networks driven by internal colored noise. The noise enters the large-$N$ theory only as a source in the self-consistent equation for the preactivation correlator,
\be
\left(-\tau^2\partial_t^2+1\right)C_\phi(t)-\lambda^2 C_f(t)=G_\xi(t)\,,
\ee
Eq.~\eqref{eq:selfrepeat}, which interpolates between white noise ($\tau_n\to0$) and quenched noise ($\tau_n\to\infty$). A matched short-/long-time solution of this equation reproduces the simulated correlators without measured input. From the same correlators we obtained two distinct dynamical thresholds: local instability at $\lambda_{c,1}$, and chaos at $\lambda_{c,2}$, determined by the ground state of a Schr\"odinger problem built from the correlator of $f'(\phi)$. Between them lies a locally unstable but non-chaotic regime, whose width shrinks as the activation approaches the linear case and as the noise becomes slower. Near the edge of chaos, weak internal noise of any finite correlation time cuts off the divergence of the intrinsic correlation time with the critical exponent, $\tau_{\rm corr,max}\sim D_0^{-1/2}$, where $D_0$ is the integrated noise strength. For weak noise the maximum sits at $\lambda_{c,2}$. Finally, the zero-frequency participation dimension is controlled by this same time scale, $D_{PR}(0)/N=(\tau/\tau_{\rm corr})^4$, so that it has a minimum scaling as $D_0^2$ at the maximum of $\tau_{\rm corr}$.

The universality of the exponent has a simple origin: close to the transition the network's own slow time $\tau_{\rm corr}$ exceeds the noise correlation time, and the noise then acts only through $D_0=\tfrac12\int G_\xi$. The exponent therefore does not depend on the shape of the noise correlations; it is a property of the network. Slow noise matters only through the crossover scale, $|f^{(3)}(0)|D_0\tau_n^2\sim\tau^3$, beyond which one approaches the static-noise case, where the divergence is not cut off at all and the two thresholds coincide. A maximal correlation time at the chaos threshold identifies it as the point of longest intrinsic memory, in line with the optimal operation reported for white noise in the locally unstable, non-chaotic window \cite{Schuecker_2018}. Our results indicate that this is not a peculiarity of white noise, although the maximum moves above $\lambda_{c,2}$ and eventually disappears as the noise strength increases.

For white noise, the frequency-dependent participation dimension and its dependence on an effective coupling were obtained in \cite{shen2025covariance}, and the quenched-noise limit was treated in \cite{Mato:2025yof}. Here we extended this to colored noise of arbitrary correlation time and we related the dimension to the intrinsic correlation time and its noise-induced cutoff, $D_{PR}(0)/N=(\tau/\tau_{\rm corr})^4$. Reduced participation has been linked to strong recurrent coupling and proximity to linear instability \cite{dahmen2020strong}, and chaotic activity can have extensive dimension occupying a small fraction of the degrees of freedom \cite{clark2023dimension}. Our model separates the relevant criteria: the participation dimension depends on $\lambda\langle f'\rangle_q$, local instability on $\lambda\langle (f')^2\rangle_q^{1/2}$, and chaos on the temporal correlations of $f'$. A small participation dimension, or an extremum of a dimensionality measure, therefore need not coincide with either threshold.

Activation shape influences both dynamical transitions and population geometry. Ref.~\cite{harish2015asynchronous} compared sigmoidal and threshold power-law responses in inhibitory rate networks, finding different chaos thresholds and fluctuation scalings near onset; for threshold power laws, the critical coupling varies nonmonotonically with the exponent and vanishes as it approaches $1/2$ from above. Their exponent describes behavior above a hard threshold, whereas our $p$ controls the large-input growth of a smooth response. Earlier correlation-statistics calculations \cite{Mato:2025yof} showed that sublinear power-law and Pad\'e-type activations regularize the linear-model variance divergence, with relative participation decreasing with the growth exponent at strong coupling, while Ref.~\cite{shen2025covariance} considered hyperbolic-tangent responses without systematically varying activation shape. In this work we examined how smooth responses with different asymptotic growth interact with the noise correlation time to control local instability, chaos, and participation. In particular, we found that the exponent $p$ controls the gap between the two thresholds, while the curvature $|f^{(3)}(0)|$ of the activation function sets the amplitude of the correlation time.

To end, let us discuss some future directions. Our analysis assumes smooth odd activation functions and a symmetric stationary state with vanishing mean activity. Extending the theory to non-odd activations would require accounting for nonzero means and distinguishing connected correlations from persistent components of the activity. The theory describes the leading large-$N$ limit, and Appendix~\ref{app:finiteN} shows appreciable size and sample dependence of the population geometry, particularly for activations close to linear. In this direction, the recent work \cite{wakhloo2026solutionchaoticneuralnetwork} was able to derive large $N$ neural correlators for fixed $W$ disorder realizations.  It would be interesting to analyze these effects in more detail.

Our dense, independent Gaussian connectivity lacks separate excitatory and inhibitory populations and outgoing-sign constraints. Experiments instead show broad, asymmetric excitatory connection strengths compatible with lognormal distributions, correlated reciprocal strengths, and overrepresented strongly connected motifs \cite{song2005highly}. Extending the model to these heterogeneous, structured ensembles is therefore important. Finally, the participation dimension characterizes the distribution of covariance eigenvalues rather than the full geometry of the activity distribution or the dimension of a dynamical attractor. The scale dependent measures proposed by \cite{recanatesi2022scale} may provide useful tools for this purpose.

\section*{Acknowledgments} 
GM is supported by CONICET (PIP grant 112202101 00114CO), CNEA and Instituto Balseiro, Universidad Nacional de Cuyo.
FR is supported by a CONICET PhD fellowship and by Instituto Balseiro, Universidad Nacional de Cuyo.
GT is supported by CONICET (PIP grant 11220200101008CO), CNEA, Instituto Balseiro, Universidad Nacional de Cuyo, and a Simons Foundation targeted grant to institutions.

\appendix

\section{Collective-field derivation of the self-consistent equation}\label{app:collective}

The self-consistent equation \eqref{eq:selfrepeat} can be derived more formally using collective fields, which also makes transparent the large-$N$ framework used in Appendix~\ref{app:fourpoint}. Starting from \eqref{eq:action1model}, we write
\be
Z= \int D\phi\,D\eta \,D\rho \,[\Det(2\pi M^{-1})]^{N/2}\,e^{-\frac{i}{2} \int_{t,t'}\eta(t,t') (\sum_j f_j(t) f_j(t')- N \rho(t, t'))} e^{-\frac{1}{2} \int_{t,t'} \sum_i  \phi_i(t) G^{-1}(t,t') \phi_i(t')}\,
\ee
where now
\be\label{eq:G2app}
(\tau \partial_t+1)(\tau \partial_{t'}+1) G(t,t') = G_{\xi}(t-t')+\lambda^2 \rho(t,t')\,.
\ee
Up to a constant, the path integral over $\phi$ gives
\be
Z= \int D\eta \,D\rho\,e^{\frac{i}{2} N \int_{t,t'}\eta(t,t') \rho(t, t')} \langle e^{-\frac{i}{2} \int_{t,t'}\eta(t,t') \sum_j f_j(t) f_j(t')} \rangle_G\,.
\ee
Since $\eta$ is self-consistently vanishing, we approximate
\be
 \langle e^{-\frac{i}{2} \int_{t,t'}\eta(t,t') \sum_j f_j(t) f_j(t')} \rangle_G \approx e^{-\frac{i}{2} N \int_{t,t'}\eta(t,t') \langle f(t) f(t') \rangle_G}\,,
\ee
so that extremizing with respect to $\eta$ recovers the self-consistent equation (\ref{eq:self1model}).

\section{Replica derivation of the four-point function}\label{app:fourpoint}

As discussed in \cite{Mato:2025yof}, computing $W$ averages of $\xi$-averaged quantities, such as \eqref{eq:Gammadefmodel} , requires introducing replicas of the neural and collective fields.
We now give the full derivation of the result quoted in Sec.~\ref{subsec:dprresult}. Adding a source term for the output,
\be
e^{\int_{t,t'} \sum_{i,a,b} J_i^{ab}(t, t') f\left(\phi_i^a(t)\right) f\left(\phi_i^b(t')\right)}\,,
\ee
symmetric in replica space and depending only on $t-t'$, our goal is to compute $\log Z$ to quadratic order in $J$. Then $\Gamma^f(\omega)$ in \eqref{eq:Gammadefmodel} is extracted as
\begin{align}\label{eq:Gammadefapp}
\Gamma^f(\omega)&=\int_{t_1,t_2} e^{i  \omega(t_1-t_2)} \langle f_i^a(t_1) f_i^b(0) f_j^b(t_2) f_j^a(0) \rangle\Big|_{i\ne j}= \frac{1}{Z} \frac{\delta}{\delta J_i^{ab}(\omega)}\frac{\delta}{\delta J_j^{ab}(-\omega)} Z\Big|_{J=0}\,.
\end{align}
Using the multi-index notation $\phi_A \equiv \phi_a(t)$, $\rho_{AB}\equiv \rho_{ab}(t,t')$, $J_{AB}\equiv\sum_j J_j^{ab}(t,t')$, and $\sum_{AB} \eta_{AB} \rho_{AB} \equiv \sum_{a,b} \int_{t,t'} \eta_{ab}(t,t') \rho_{ab}(t,t')$, the replicated collective-field path integral is
\be
Z= \int D\eta \,D\rho\,e^{N \sum_{AB}\left(\frac{i}{2} \eta_{AB}+ \frac{1}{N} J_{AB}\right) \rho_{AB}} \langle e^{ -\frac{i}{2} \sum_{AB}  \eta_{AB} f_A f_B } \rangle_G^N=\int D\eta\,D\rho\,e^{-N S_{eff}}\,,
\ee
with
\be
S_{eff}=-\sum_{AB}\left(\frac{i}{2} \eta_{AB}+ \frac{1}{N} J_{AB}\right) \rho_{AB}-  \log \langle e^{-S_{site}} \rangle_G\,,\qquad
S_{site}=\frac{i}{2} \sum_{AB}\eta_{AB} f_A f_B \,.
\ee
The single-site Green's function is related to $\rho$ by 
\be
(\tau\partial_t+1)(\tau\partial_{t'}+1)G_{ab}(t,t')=G_\xi(t-t')\delta_{ab}+\lambda^2\rho_{ab}(t,t')\,,
\ee 
or, with $U_A\equiv \tau\partial_t+1$, 
\be
\rho_{AB}=\lambda^{-2}[U_AU_BG_{AB}-G^\xi_{AB}]\,.
\ee

\subsection{Saddle-point equations}

Extremizing $S_{eff}$ with respect to $\eta$ and $G$ gives
\be\label{eq:Gsaddlereplicaapp}
U_A U_B G_{AB} =G^{\xi}_{AB}+\lambda^2  \frac{1}{\langle e^{-S_{site}} \rangle_G} \langle f_A f_B e^{-S_{site}} \rangle_G\,,
\ee
which generalizes \eqref{eq:self1model} to the replicated case. We will use the Gaussian-average identity (Price's theorem)
\be\label{eq:parGapp}
\frac{\delta }{\delta G_{AB}} \langle F(\phi)\rangle_G = \frac{1}{2} \left\langle \frac{\delta^2 F(\phi)}{\delta \phi_A \delta \phi_B}\right\rangle_G\,\,.
\ee

We expand
\be
G_{AB}=G^{(0)}\delta_{AB}+N^{-1}G^{(1)}_{AB}+\ldots\;,\;\eta_{AB}=\eta^{(0)}+N^{-1}\eta^{(1)}_{AB}+\ldots\,.
\ee
At leading order we find  $\eta^{(0)}=0$ and $G^{(0)}$ solves \eqref{eq:self1model}. Since $f$ is odd, $\langle f\rangle=0$ and the off-diagonal components of $G_{AB}$ are $O(1/N)$, so we only need external sources $J_{AB}$ with vanishing diagonal. The $O(1/N)$ piece of \eqref{eq:Gsaddlereplicaapp} becomes, after using $[G^{(0)}]^{-1}_{A_1A_2}\langle f_A\phi_{A_1}\rangle_{G^{(0)}}=\delta_{A_2A}\langle f'_A\rangle_{G^{(0)}}$,
\be\label{eq:G1newapp}
U_A U_B G^{(1)}_{AB} =\lambda^2  \Big[- i\sum_{CD} \eta^{(1)}_{CD} \langle f_A f_C \rangle_{G^{(0)}} \langle f_B f_D \rangle_{G^{(0)}}+  G^{(1)}_{A B}  \langle f'_{A} \rangle_{G^{(0)}} \langle f'_B \rangle_{G^{(0)}}\Big]\,,
\ee
while the analogous $O(1/N)$ piece of the $\eta$-saddle equation fixes $\eta^{(1)}$ in terms of the source,
\be\label{eq:etanew2app}
\left[ U_A^\dag U_B^\dag-\lambda^2 \langle f'_{A} \rangle_{G^{(0)}} \langle f'_B \rangle_{G^{(0)}} \right]\eta_{AB}^{(1)}= 2i U_A^\dag U_B^\dag J_{AB}\,,\qquad U_A^\dag\equiv -\tau\partial_t+1\,.
\ee
Combining, and using time-translation invariance ($J_{ab}(t,t')=J_{ab}(t-t')$, so $\eta^{(1)}, G^{(1)}$ depend only on $t-t'$), the Fourier transforms of \eqref{eq:G1newapp}-\eqref{eq:etanew2app} give, with $\lambda_{eff}=\lambda\langle f'(0)\rangle_{G^{(0)}}$,
\be\label{eq:etanew4app}
\eta_{ab}^{(1)}(\omega)= 2i \frac{1+\tau^2 \omega^2}{1+\tau^2 \omega^2-\lambda_{eff}^2 } J_{ab}(\omega)\,,\qquad
G^{(1)}_{ab}(\omega) =2 \lambda^2 \frac{1+\tau^2 \omega^2}{(1+\tau^2 \omega^2-\lambda_{eff}^2 )^2}C_f(\omega) C_f(-\omega) J_{ab}(\omega)\,.
\ee

\subsection{On-shell effective action and $\Gamma^f$}

To evaluate $\log Z$ to quadratic order in the site-dependent (not just collective) sources $J_{i,AB}$, we decompose $J_{i,AB}=N^{-1}J_{AB}+\delta J_{i,AB}$ with $\sum_i \delta J_{i,AB}=0$; the collective saddle depends only on $J_{AB}=\sum_i J_{i,AB}$, but the quadratic generating functional also contains a local piece $\propto \sum_i J_iJ_i$, which must be kept to correctly isolate the coefficient of the collective piece. Expanding the single-site cumulant generating function to second order and using the saddle-point equations \eqref{eq:etanew2app} to simplify the collective terms, the on-shell quadratic effective action simplifies to
\be\label{eq:SeffOScompactapp}
S_{eff,on}^{(2)}=-\frac{1}{N}\sum_i\delta J_i\,\mathcal K\,\delta J_i+\frac{1}{4N^2}\eta^{(1)}\,\mathcal K\,\eta^{(1)}\,,
\ee
where, for two off-diagonal bilocal kernels $B_{AB}$, $D_{AB}$, $B\,\mathcal K\,D\equiv \sum_{ABCD}B_{AB}\,C^{(0)}_{AC}C^{(0)}_{BD}\,D_{CD}$ with $C^{(0)}_{AB}=\delta_{ab}C_f(t-t')$. Using \eqref{eq:etanew4app} and writing $J_{ab}(\omega)=\sum_i J_i^{ab}(\omega)$, this becomes
\begin{align}
S_{eff,on}^{(2)}={}&-\frac{1}{N}\sum_{i,a\neq b}\int_\omega C_f(\omega)C_f(-\omega)\, J_i^{ab}(-\omega)J_i^{ab}(\omega)\nonumber\\
&-\frac{1}{N^2}\sum_{a\neq b}\int_\omega\, C_f(\omega)C_f(-\omega)\left[\frac{(1+\tau^2\omega^2)^2}{(1+\tau^2\omega^2-\lambda_{eff}^2)^2}-1\right]J_{ab}(-\omega)J_{ab}(\omega)\,.
\end{align}
where we used $\sum_i \delta J_i^{ab}(-\omega)\delta J_i^{ab}(\omega)=\sum_i J_i^{ab}(-\omega)J_i^{ab}(\omega)-N^{-1}J_{ab}(-\omega)J_{ab}(\omega)$. The first term is proportional to $\sum_i J_iJ_i$ and contributes only for $i=j$; the second, proportional to $(\sum_iJ_i)^2$, gives the desired $i\ne j$ response. Since $Z\simeq e^{-NS_{eff,on}}$ and the one-point function in the off-diagonal replica sector vanishes at $J=0$, the second derivative in \eqref{eq:Gammadefapp} picks out only the second term, giving \eqref{eq:Gammafinalmodel} and hence \eqref{eq:DPRomega}.

\section{Finite $N$ effects}\label{app:finiteN}

In this appendix we analyze numerically finite $N$ and fixed $W$-noise realization effects.

\begin{figure}[h!]
	\centering
	\includegraphics[width=0.6\textwidth]{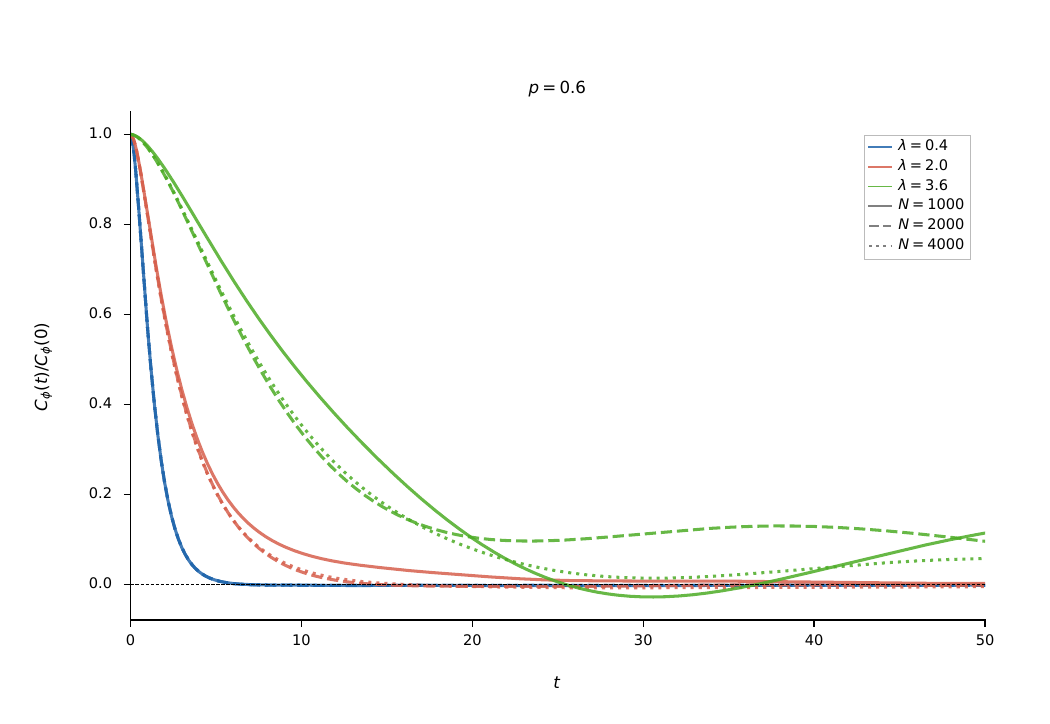}
	\caption{Normalized preactivation autocorrelation for
$N=1000,2000,4000$, at $\tau_n=0.5$, $D_\xi=40$, $p=0.6$,
and $\lambda=0.4,2.0,3.6$. Size dependence is more pronounced
at the larger couplings.}
	\label{fig:correlatorsfiniteN}
\end{figure}

We first examine finite-size effects in the temporal correlations.
Figure~\ref{fig:correlatorsfiniteN} compares the normalized input
autocorrelation for $N=1000,2000,4000$, at $p=0.6$,
$\tau_n=0.5$, and $D_\xi=40$. At $\lambda=0.4$, the curves nearly
overlap, while appreciable size dependence appears at $\lambda=2.0$
and $3.6$. For the strongest coupling, differences extend to the
long-lag part of the correlator. Since each curve is normalized by
its own equal-time value, these differences concern the correlation
profile as well as the activity variance. These effects have an intuitive origin in the non-saturation of the activation function, which allow more rare network fluctuations.

To examine sample-dependent population geometry we analyze independent activity clouds. An initial scan covered $N=2000,4000,8000$, five coupling matrices, and several couplings at $p=0.6$ and $0.8$, using a transient of duration $300$. A focused control then used $N=2000,4000,8000,12000$ and doubled the transient to $600$, at $(p,\lambda)=(0.6,5),(0.8,3),(0.8,3.5)$. Both studies used $\tau=1$, $\beta=0.6$, $D_\xi=10$, $\tau_n=2$, and $\delta t=0.1$. In these cloud simulations, the OU variable is updated
before the activity, so the latter uses $\xi_i(t+\delta t)$.
The numerical protocol of Sec.~\ref{subsec:numerics} instead
uses $\xi_i(t)$. We have checked the sensitivity to this update
order numerically and found it negligible for the comparisons
considered here. The focused control follows a common raw disorder sample as the network grows. For each seed, a matrix $G$ of size $N_{\max}=12000$ is drawn with independent unit-variance Gaussian entries, and the physical couplings at size $N$ are
\begin{equation}\label{eq:finiteNnestedW}
W^{(N)}_{ij}=\frac{\lambda}{\sqrt{N}}G_{ij},\qquad i,j\le N.
\end{equation}
Thus the common block has the same raw entries across sizes, while the normalization changes to preserve Eq.~\eqref{eq:Wstats}. This is a correlated sequence of networks from the prescribed ensemble, rather than an unchanged small matrix with extra neurons appended.

\begin{figure}[h]
\centering
\includegraphics[width=0.7\textwidth]{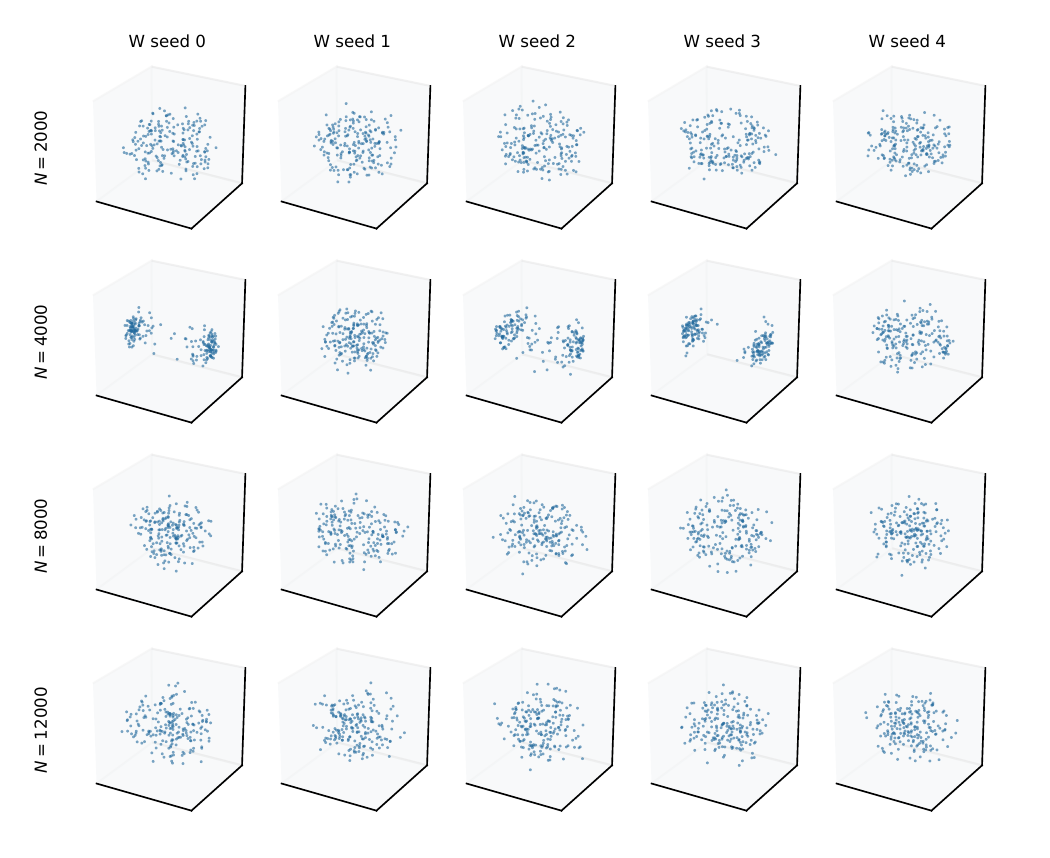}
\caption{Activity clouds for $p=0.6$, $\lambda=5$, under nested disorder. Rows correspond to $N=2000,4000,8000,12000$; columns to five independent raw matrices $G$. Each panel projects $200$ final states onto that network's first three sample PCs. Coordinates are divided by $\sqrt{N}$, with a common display scale within each row. The visibly separated two-lobe clouds at $N=4000$ are absent in the displayed $N=8000$ and $12000$ samples.}
\label{fig:finiteNcloud06}
\end{figure}

\begin{figure}[h]
\centering
\includegraphics[width=0.6\textwidth]{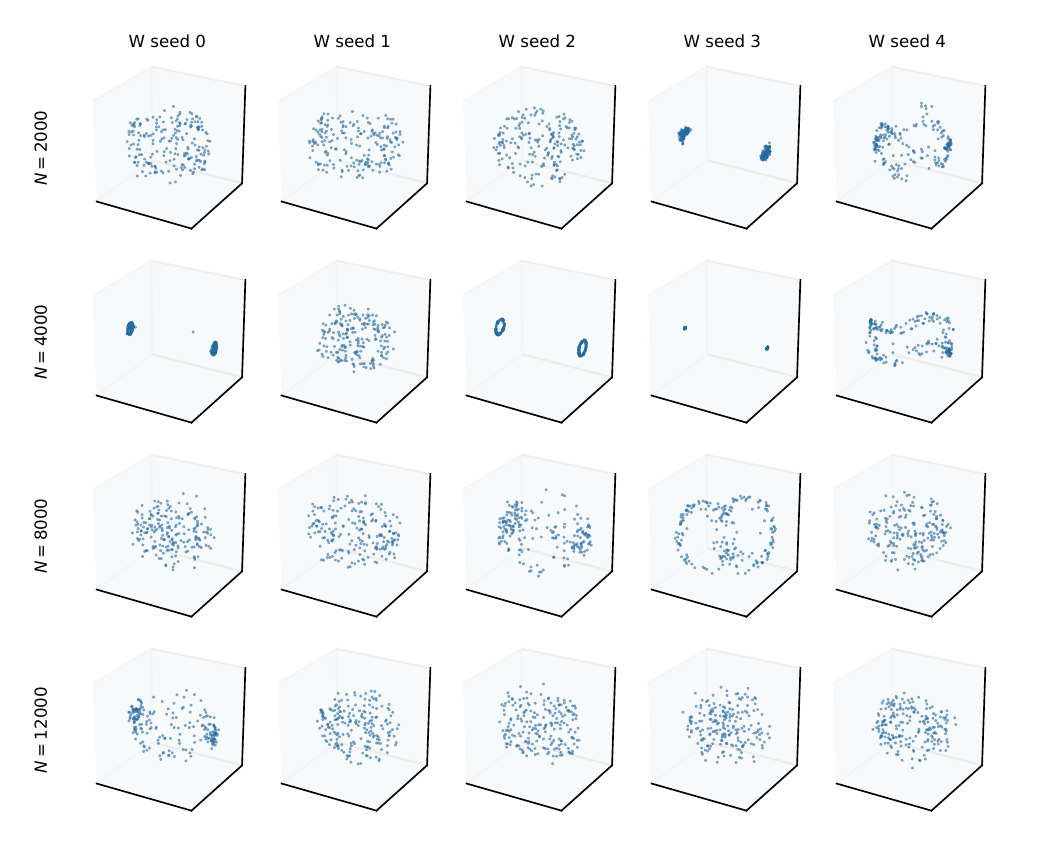}
\caption{Nested-disorder activity clouds for $p=0.8$, $\lambda=3.5$, with the layout and sampling protocol of Fig.~\ref{fig:finiteNcloud06}. Two compact regions occur for some small and intermediate networks, while larger networks retain sample-dependent broad or annular projections. Display scales are common within a row; row-to-row visual sizes are not amplitude measurements.}
\label{fig:finiteNcloud08}
\end{figure}

Each cloud contains one final state from each of $200$ independent initial-condition and noise realizations at fixed $W$, after the prescribed transient. Principal Component Analysis (PCA) is applied to the centered sample cloud separately for each network. The leading-PC variance fraction $F_1$ is the proportion of the total cloud variance captured by PC1, the direction of greatest spread. Values close to one indicate that a single population direction accounts for most of the variance; this measure does not by itself establish bimodality. The PCA coordinates are descriptive projections, not consecutive points on a trajectory. Figures~\ref{fig:finiteNcloud06} and \ref{fig:finiteNcloud08} show the first three coordinates. We also retain the signed first coordinate $z={\rm PC1}/\sqrt{N}$ when inspecting individual clouds, and the sign-invariant variable $|z|$ for summaries over $W$. The normalization removes the explicit $\sqrt{N}$ scale of a coherent population projection; it does not by itself establish how the projection scales with size. Distinct networks have distinct PCA bases, so their three-dimensional coordinates are not averaged.

At $p=0.6$, $\lambda=5$, the $N=2000$ projections are relatively diffuse. At $N=4000$, seeds $0$, $2$, and $3$ display two separated lobes, whereas the other samples remain more broadly distributed. This structure is no longer clearly separated at $N=8000$ or $12000$. The leading-PC variance fraction confirms the nonmonotonic size dependence. For seed $0$, it changes from $0.194$ to $0.609$, $0.081$, and $0.096$ as $N$ increases; the five-sample mean is $0.185$, $0.389$, $0.111$, and $0.083$. Hence increasing $N$ first strengthens and then weakens the concentration along the leading sample PC. The lobes at $N=4000$ cannot be extrapolated as a persistent large-$N$ property from that size alone.

\begin{figure}[h]
\centering
\includegraphics[width=0.7\textwidth]{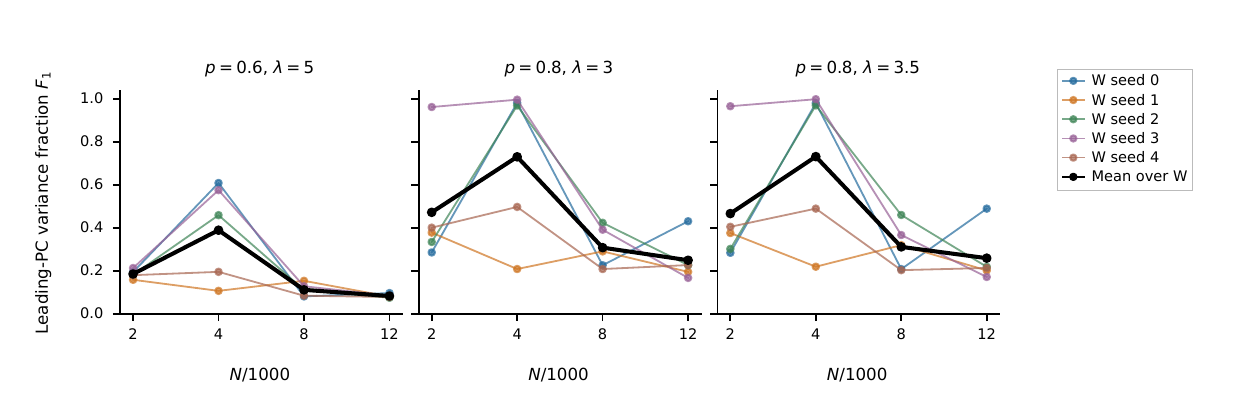}
\caption{Fraction of the centered cloud variance explained by its leading sample PC, $F_1=\ell_1/\sum_k\ell_k$, for the three nested-disorder cases. Each thin colored line follows one raw disorder seed across $N$; the thick black line is their arithmetic mean. Each fraction is computed from 200 independent simulated final
states per network. The prominent intermediate-size concentration is followed by a reduction at larger $N$, with more persistent spread for $p=0.8$. This equal-time sample diagnostic is distinct from the spectral participation fraction.}
\label{fig:finiteNpc1}
\end{figure}

For $p=0.8$, both $\lambda=3$ and $3.5$ produce more pronounced sample dependence. Several $N=4000$ clouds are concentrated near two compact regions, while other samples are diffuse or approximately annular. At larger sizes the compact two-region pattern is replaced by broader clouds in the samples studied, although strong anisotropy and annular projections can remain. At $\lambda=3.5$, the leading-PC fractions at $N=4000$ range from $0.219$ to $0.999$ across $W$. Their mean decreases from $0.732$ to $0.311$ and $0.259$ at $N=8000$ and $12000$, but the last set still ranges from $0.171$ to $0.490$. Figure~\ref{fig:finiteNpc1} shows the individual fractions for all three focused cases. This residual spread prevents a claim that all population geometry has converged by $N=12000$.

\begin{figure}[h!]
\centering
\includegraphics[width=0.7\textwidth]{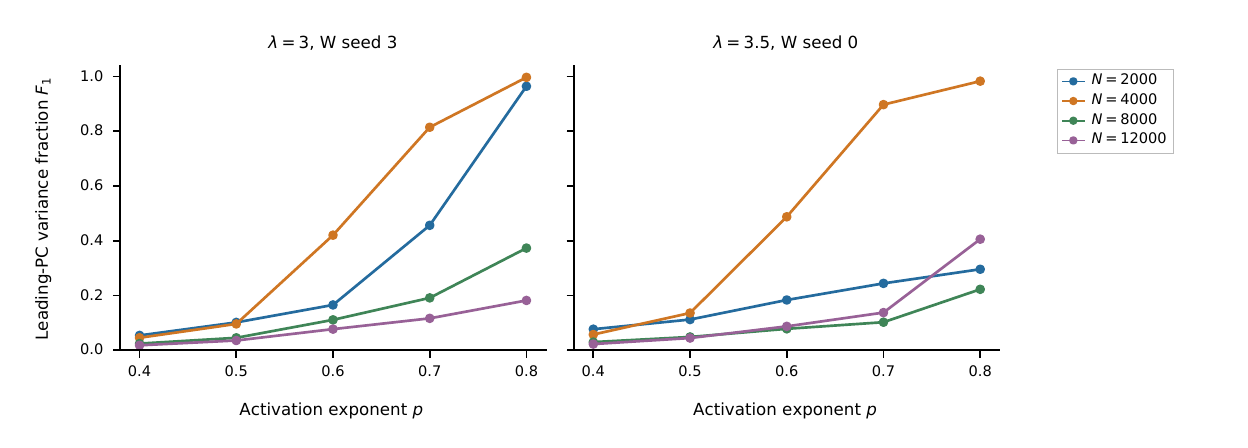}
\caption{Leading-PC variance fraction of the input clouds versus activation exponent, at fixed $(\lambda,W\text{-seed})=(3,3)$ (left) and $(3.5,0)$ (right), for four nested sizes. Each fraction is computed from $200$ independent final states after a transient of duration $600$. The matrices are nested across $N$, and the noise sequences are paired across $p$ at fixed size and pair. Lines connect measured values without locating a transition between them.}
\label{fig:finiteNpcontrol}
\end{figure}

A further scan holds $(\lambda,W\text{-seed})$ fixed at $(3,3)$ or $(3.5,0)$, varies $p=0.4,0.5,0.6,0.7,0.8$, and uses the same four nested sizes and a transient of duration $600$. At fixed size and pair, the simulations across $p$ reuse the same pseudorandom sequences for initial conditions and OU noise. The leading-PC fractions increase strongly with $p$, especially at $N=4000$, as shown in Fig.~\ref{fig:finiteNpcontrol}. For $(\lambda,W\text{-seed})=(3,3)$ at this size, the input-cloud fraction rises from $0.045$ at $p=0.4$ to $0.997$ at $p=0.8$; at $N=12000$ the corresponding fractions are $0.018$ and $0.182$. This scan reinforces the dependence on both activation and size, but its two selected samples do not determine a disorder-averaged boundary in $p$.

These observations establish appreciable finite-size and disorder-sample dependence of projected population activity. They do not establish failure of self-averaging in the infinite-size limit: that would require a controlled asymptotic analysis of a specified observable and its disorder variance. Likewise, two lobes do not prove distinct attractors, and an annular snapshot cloud does not prove a periodic orbit or a Hopf bifurcation. The longer transient provides a relaxation control, but the snapshots contain no switching times or time ordering. Finally, the rank of a centered cloud covariance estimated from $200$ states is at most $199$; its leading-PC fractions are sample diagnostics rather than unbiased estimates of the full covariance spectrum. The present evidence therefore motivates a separate study of these sample-dependent collective structures, while supporting the more limited finite-size conclusions stated here.

\newpage
\bibliography{RNN}{}
\bibliographystyle{utphys}

\end{document}